\documentclass[a4paper,fleqn]{cas-sc}

\usepackage[numbers,sort&compress]{natbib}
\usepackage{lineno,hyperref}
\modulolinenumbers[5]
\usepackage{xfrac}
\usepackage{graphicx}
\usepackage{subfig}
\graphicspath{ {./figures/} }
\usepackage{fullpage}
\usepackage{float}
\usepackage{caption}
\usepackage{siunitx}
\usepackage{bm}

\usepackage{xcolor}
\usepackage{mathtools}
\usepackage{xurl}
\usepackage{bigfoot}
\DeclareNewFootnote[para]{default}

\newcommand{\pf}{\varphi}
\newcommand{\gs}{g_s}
\newcommand{\conc}{u}
\newcommand{\temp}{T}
\newcommand{\Dt}{\Delta t}
\newcommand{\pfh}{\pf_h}
\newcommand{\conch}{\conc_h}
\newcommand{\temph}{\temp_h}
\newcommand{\thetam}{\theta}
\newcommand{\atheta}{\vartheta}

\newcommand{\jfeps}{\varepsilon}
\newcommand{\jat}{{\bf j}_{\mbox{at}}}

\newcommand{\bU}{{\bf U}}

\newcommand{\bx}{{\bf x}}

\newcommand{\bM}{{\bf M}}
\newcommand{\bv}{{\bf v}}

\newcommand{\Jac}{\mathcal{J}}

\newcommand{\Of}{\Omega}
\newcommand{\R}{\mathbb{R}}

\newcommand{\beqn}{\begin{equation}}
\newcommand{\eeqn}{\end{equation}}

\newcommand{\pd}[2]{\frac{\partial#1}{\partial#2}}
\newcommand{\dif}[1]{\,\mathrm{d}\-#1}

\newcommand{\dt}{\Delta t}

\newcommand{\tusas}{\texttt{Tusas}}
\newcommand{\summit}{\texttt{Summit}}
\newcommand{\sierra}{\texttt{Sierra}}

\renewcommand{\Jac}{\mathbf{F'}}
\renewcommand{\Prec}{\mathbf{M}}

\begin{document}
\let\WriteBookmarks\relax
\def\floatpagepagefraction{1}
\def\textpagefraction{.001}
\shortauthors{S. Ghosh et~al.}
%\begin{frontmatter}

\title [mode = title]{\tusas : A fully implicit parallel approach for coupled phase-field equations}                    

% Group authors per affiliation:
\author[1]{Supriyo Ghosh}\cormark[1]
\author[1]{Christopher K. Newman}
\author[2]{Marianne M. Francois}
\cortext[cor1]{\mbox{Corresponding author. \textit{Email address:} \texttt{gsupriyo2004@gmail.com}} (Supriyo Ghosh)}
\address[1]{Fluid Dynamics and Solid Mechanics Group (T-3), Los Alamos National Laboratory, Los Alamos, NM 87545, USA}
\address[2]{Theoretical Division (T-DO), Los Alamos National Laboratory, Los Alamos, NM 87545, USA}

%[orcid=0000-0001-7265-5266]

\begin{abstract}
We develop a fully-coupled, fully-implicit approach for phase-field modeling of solidification in metals and alloys. Predictive simulation of solidification in pure metals and metal alloys remains a significant challenge in the field of materials science, as microstructure formation during the solidification process plays a critical role in the properties and performance of the solid material. Our simulation approach consists of a finite element spatial discretization of the fully-coupled nonlinear system of partial differential equations at the microscale, which is treated implicitly in time with a preconditioned Jacobian-free Newton-Krylov method. The approach %allows timesteps larger than those restricted by the traditional explicit CFL limit and 
is algorithmically scalable as well as efficient due to an effective preconditioning strategy based on algebraic multigrid and block factorization. We implement this approach in the open-source \tusas\ framework, which is a general, flexible tool developed in C++ for solving coupled systems of nonlinear partial differential equations. The performance of our approach is analyzed in terms of algorithmic scalability and efficiency, while the computational performance of \tusas\ is presented in terms of parallel scalability and efficiency on emerging heterogeneous architectures. We demonstrate that modern algorithms, discretizations, and computational science, and heterogeneous hardware provide a robust route for predictive phase-field simulation of microstructure evolution during additive manufacturing.
\end{abstract}


\maketitle

%-----------------------------------------------------------------
\section{Introduction}\label{sec:intro}
Microstructure evolution during the processing stage determines the properties of the final material in manufacturing applications. The key materials processes are categorized broadly as solidification and solid-state phase transformations. Solidification controls the size, shape, distribution, and morphology of the microstructure grains, and the subsequent solid-state phase transformation processes control the grain growth. Therefore, understanding of the solidification behavior of materials is essential and often regarded as a benchmark test problem for verification and validation of concerned numerical approaches. In particular, dendrite growth during solidification of metal materials is commonly studied and utilized for verification and validation of solidification models~\cite{dantzigbook,kurzbook}. 

Recently there has been an identified need for an improved predictive simulation capability for additive manufacturing (AM) processes~\cite{francois2017,ghosh_review}. Improved simulation capability will allow many important AM-specific physics questions to be answered for the first time and is necessary to improve quality, decrease processing time, and eventually minimize costs of an AM product. An invaluable, individual component to this capability is the ability to simulate microstructure evolution during solidification. The phase-field approach has demonstrated particular success for microstructure evolution during the rapid solidification in metal AM processes, resulting in dendritic microstructures~\cite{Trevor2017,ji2018_pfam,ranadip2017,Kundin2015,ghosh_msmse2017,ghosh2018_jom,ghosh2018_ti64,karayagiz2020}.

The phase-field method~\cite{chen2002,boettinger2002,moelans2008,steinbach2009} is a common approach for quantitative modeling of dendrite growth. Depending on the nature of the process, the phase-field method addresses multiple variables, including composition, anisotropy, phase, temperature, orientation, and stress, fully-coupled as a set of nonlinear partial differential equations (PDEs), typically posed on high-resolution meshes. There are challenges to applying phase-field methods for predictive simulation of solidification microstructures. Three critical challenges include:
\begin{enumerate}
\item 
Ability to control restrictive time scales due to stability constraints and stiff physics.
\item 
Ability to control restrictive length scales and inherent large problem size due to mesh resolution necessary to resolve the diffuse interface and a large set of fully-coupled nonlinear PDEs.
\item 
Code flexibility required to couple a large set of arbitrary PDEs and multiphysics at run time, while effectively and efficiently leveraging modern computer architectures and software.
\end{enumerate}
Many current phase-field simulation efforts utilize explicit time integration methods, which can be impractical due to constraints on the maximum stable timestep due to the Courant-Friedrichs-Lewy (CFL) condition~\cite{lambert:1991}. Additionally, many efforts utilize finite difference or finite volume spatial discretizations that are limited to second-order accuracy on regular, structured meshes. Furthermore, many phase-field codes are implemented using FORTRAN, which poses difficulties in utilizing modern libraries and to leverage modern architectures effectively.

Our approach to phase-field modeling particularly addresses the above challenges. We mitigate time scale challenges by utilizing implicit temporal integration, which allows timesteps larger than the explicit CFL limit and readily affords second-order and higher time integration. A direct advantage of implicit methods is that relatively large, second-order timesteps can be taken, since the timestep is determined by accuracy instead of by numerical stability constraints. In other words, timestep sizes can be chosen such that a \emph{desired level of accuracy} can be attained independent of stability conditions. A disadvantage, however, is that fully implicit methods require a nonlinear solution be attained for each timestep. 

Additionally, we utilize finite element spatial discretization methods~\cite{brenner2007fem,braess01} for mitigation of length scale challenges. The finite element method allows second-order and third-order approximations on unstructured meshes, and is easily implemented in three-dimensions (3D), and is particularly suited for simulation of complex geometries such as solidification melt pools in AM processes. 

There are two different approaches for solving the system of coupled nonlinear PDEs. In a monolithic approach~\cite{monolithic1,monolithic2}, the entire fully-coupled nonlinear system of equations is solved within a single time increment; whereas, a partitioned approach~\cite{partitioned1,partitioned2,joshi2018adaptive,joshi2018positivity} divides the coupled system into an iteration of several subproblems so that individual fields are treated successively. We use a monolithic approach, which solves the multiphysics system within a single finite element program and is unconditionally stable for implicit time integration schemes. However, for strongly coupled multiphysics and multiscale problems, such as electro-thermo-fluid-mechanical interactions in materials, a partitioned approach can be suitable and enables the use of different kinds of software, solvers, and discretizations for the fields concerned.

In recent years, there has been considerable advancement in both linear and nonlinear solution approaches to multiphysics applications, in particular Jacobian-free Newton-Krylov (JFNK) methods~\cite{knoll:2004,kelley95,knoll:2000a,saad95,joshi2018positivity,joshi2019hybrid,joshi2018adaptive}. 
The key to efficient implementation of JFNK is effective preconditioning. In particular, physics-based preconditioning~\cite{knoll:2000a,knoll:2005,park:2010} has been successful. This unique preconditioning approach is a highly successful and proven approach for effective preconditioning for multiple time scale problems, where an accurate simulation is desired on the dynamical time scale. Effective preconditioning reduces the number of linear solver iterations per Newton iteration and allows larger timesteps by \emph{stepping over} stiff physics common to elliptic operators. Thus, the timestep is determined by accuracy instead of by numerical stability constraints~\cite{knoll:2004,mousseau:2002}. Furthermore, gains in efficiency and accuracy have been understood through a modified equation analysis of splitting and linearization errors~\cite{knoll:2003}. These algorithms, combined with modern software and libraries, provide a robust route to flexible multiphysics simulations on modern, emerging heterogeneous supercomputer architectures.

The phase-field approach is computationally intensive; a high computational cost is required to resolve scientifically relevant length and time scales associated with microstructure patterns that result due to phase change, species diffusion, and heat flow during solidification. To accurately simulate such multiphysics phenomena, it is often necessary to perform large-scale simulations at small timesteps, requiring many hours or days of conventional computational resources. Parallel computing approaches \textit{\textit{via}} distributed memory Message Passing Interface (MPI)~\cite{Mpich}, shared memory OpenMP~\cite{Openmp}, and Compute Unified Device Architecture (CUDA) for Graphics Processing Units (GPUs)~\cite{Cuda} are currently being utilized to attain computational performance and parallel scalability. Recently, hybrid approaches \textit{via} MPI+OpenMP, for CPU-parallel architectures, or MPI+CUDA, for CPU/GPU architectures, have been adopted to further speed up the simulations in a variety of applications~\cite{jacobsen2010hybrid,glaser2015hybrid,komatitsch2010hybrid}. The advantages of hybrid MPI+CUDA-based parallel computing in which both computational power and wide memory bandwidth will allow large-scale phase-field simulations to achieve superior performance and parallel scalability~\cite{yang2017gpu, yamanaka2011gpu,sakane2015gpu}.

Software development for phase-field simulation is an emerging need for multiphysics materials modeling capability. Existing software within the phase-field community include \texttt{MOOSE}~\cite{moose}, \texttt{PACE3D}~\cite{pace3d}, \texttt{MMSP}~\cite{mmsp}, \texttt{MICRESS}~\cite{micress}, \texttt{OpenPhase}~\cite{openphase}, \texttt{PhasePot}~\cite{phasepot}, \texttt{FiPy}~\cite{fipy}, \texttt{PRISMS-PF}~\cite{stephen2020}, \texttt{AMPE}~\cite{dorr2010ampe}, and \texttt{MEUMAPPS}~\cite{rad2018}. These frameworks, each containing phase-field models targeted for different applications with different sets of coupled nonlinear PDEs, are implemented with various discretizations and algorithmic approaches. To date, the majority of these efforts are based on finite difference spatial discretization or explicit time integration. Some frameworks such as \texttt{MOOSE}~\cite{moose} and \texttt{PACE3D}~\cite{pace3d} are implemented in 3D on conventional CPU-parallel architectures; however, these codes are not known to scale and do not make use of next-generation heterogeneous CPU/GPU architectures necessary to accelerate computational performance required for materials simulation and development. Currently, there are only a few computational efforts on phase-field methods implemented within hybrid CPU/GPU architectures~\cite{zhu2018gpu,yamanaka2011gpu,shimokawabe2011}. However, these are single-purpose codes, with numerical approaches limited only to specific phase-field implementations and thus can not be easily generalized to incorporate additional multiphysics modules.

In this work, we implement phase-field methods for solidification within the \tusas\ software environment. \tusas\ is a general, flexible code for solving coupled systems of nonlinear PDEs and was originally developed particularly for phase-field simulation of solidification. The \tusas\ approach consists of a finite element spatial discretization of the fully-coupled nonlinear system, which is treated explicitly or implicitly in time with a preconditioned JFNK method. In addition, \tusas\ leverages modern computational science and state-of-the-art software libraries that enable scalable and efficient predictive simulation of microstructure evolution on heterogeneous architectures.

This manuscript is organized as follows: a description of the phase-field equations, finite element and temporal discretizations, and solution strategies are presented in Sec.~\ref{sec:form}; an overview of \tusas\ is outlined in Sec.~\ref{sec:features}; results in terms of accuracy, algorithmic efficiency, and parallel performance for two academic problems: solidification of homogeneous materials and solidification of dilute binary alloys, in addition to a comparison to explicit temporal methods, are presented in Sec.~\ref{sec:results}; conclusions are drawn in Sec.~\ref{sec:summary}. 

%---------------------------------------------------------------------
\section{Mathematical approach}\label{sec:form}
The phase-field method~\cite{chen2002,boettinger2002,steinbach2009,moelans2008} is a powerful technique utilized to simulate microstructure evolution during solidification. A particular strength of the method is that the sharp solid-liquid interface is modeled by a well-localized, but diffuse interface with a finite width; which reduces to a set of nonlinear PDEs that are handled by standard numerical techniques. As we do not seek to detail the phase-field technique here, we only describe the model equations to convey the numerical schemes adopted in \tusas.

\subsection{Problem description: Phase-field technique}\label{sec:model}
Let the scalar variable phase-field $\pf(\bx, t)$, solute concentration $\conc (\bx, t)$, and temperature $\temp (\bx, t)$, describe the non-equilibrium dynamics of a material undergoing the solidification process, resulting in microstructure patterns. The variable $\pf$ describes the physical state of the microstructure regions. While the bulk phases, solid and liquid, are represented by constant values of $\pf$, the solid-liquid interface is identified by a continuous transition of $\pf$ between bulk phases. With this description, the interface is not tracked explicitly but is given implicitly by a contour of $\pf$, making the implementation relatively straightforward compared to sharp-interface methods~\cite{caginalp1989,sekerka2004} and easily extendable to 3D. For multiphase systems, suitable interpolation functions of $\pf$ are used to adjust the material properties independently across bulk and interface regions. Therefore, in the diffuse-interface description, no conditions at the interface between the phases are required to be satisfied. Let the material occupy a spatial region $\Of\subset\R^3$ with boundary $\partial\Of$, with $\bx \, (\in\Of) = [x\ y\ z]^\mathrm{T}$ is the spatial variable, and $t\in [0,\infty)$ is time.
The material is described by the Helmholtz free energy functional~\cite{chen2002,boettinger2002,steinbach2009,moelans2008},
\begin{equation}\label{eq:functional}
    \mathcal{F} = \int_\Omega \left[ f(\pf,u,T) + \frac{\sigma^2}{2}|\nabla\pf|^2 \right] \, \dif{\bx},
\end{equation}
where $f(\pf,\conc,\temp)$ is the bulk free energy, and the second term is the interface free energy in which the coefficient $\sigma^2$ relates to the width, $W$, of the diffuse interface. Although Eq.~\eqref{eq:functional} is the starting point for most phase-field models, there can be additional terms due to the coupling of the field variables and contributions from driving forces during the materials process, see~\cite{chen2002,boettinger2002,steinbach2009}. The minimization of the functional $\mathcal{F}$ with respect to field variables~\cite{arfken1999,ewing:1969} yields the following PDEs for temporal evolution of field variables,
\begin{equation}\label{eq:pf}
    \pf_t +\frac{1}{\tau} \frac{\delta \mathcal{F}}{\delta \pf} = 
    \pf_t+\frac{1}{\tau} \left(\pd{f}{\pf} - \sigma^2 \nabla\cdot\nabla\pf\right)=0,
\end{equation}

\begin{equation}\label{eq:c}
     \conc_t - \nabla \cdot \left(D \nabla \frac{\delta \mathcal{F}}{\delta \conc}\right)  = \conc_t- \nabla \cdot \left(D \nabla \pd{f}{\conc}\right)=0,
\end{equation}
and
\begin{equation}\label{eq:temp}
     \temp_t - \alpha \nabla\cdot\nabla\temp - G_1 =0.
\end{equation}
Here, the subscript $t$ denotes partial differentiation with respect to $t$, $\nabla\cdot$ and $\nabla$ are the divergence and gradient operators, $\tau$ is the time scale related to atomic mobility at the interface, and $D$ is the diffusivity related to mobility of solute in the liquid. Equation~\eqref{eq:pf} is the Allen-Cahn equation~\cite{Allen1979} for the non-conserved variable $\varphi$, Eq.~\eqref{eq:c} is the Cahn-Hilliard diffusion equation~\cite{Cahn1958} for the conserved field $\conc$, and Eq.~\eqref{eq:temp} is a standard thermal diffusion equation with thermal diffusivity $\alpha$ and heat source term for solidification $G_1 = L/c_p\,\pf_t$, where $L$ is the latent heat and $c_p$ the specific heat. For the example problems in this manuscript, we employ zero Neumann boundary conditions on $\partial\Of$ and specify initial conditions for $\pf(\bx,0)$, $\conc(\bx,0)$, and $\temp(\bx,0)$.

\subsubsection{Finite element formulation}\label{sec:fem}
We utilize the weak form of Eqs.~\eqref{eq:pf}--\eqref{eq:temp} and choose an appropriate subspace of $V\subset H^1(\Omega)$ approximated by $V_h$, where $V_h$ is a Lagrange (nodal) finite element spatial discretization such that $V_h\subset V$ and $V_h=\mbox{span}\{\psi_j(\bx)\}_{j=1}^E$. Here, $h$ is a discretization parameter chosen as a measure of the mesh size, and $E$ is the number of nodes in the mesh. Let $\pfh=\sum_{j=1}^E \pf_j(t) \psi_j(\bx)$, $\conch=\sum_{j=1}^E \conc_j(t) \psi_j(\bx)$, and $\temph=\sum_{j=1}^E \temp_j(t) \psi_j(\bx)$ be the discretizations in $V\times[0,\infty)$ of $\pf$, $\conc$, and $\temp$, respectively. The finite element form of Eqs.~\eqref{eq:pf}--\eqref{eq:temp} is given by:
%find ${\pfh}_t$, ${\conch}_t$ and ${\temph}_t$ such that
\begin{equation}\label{eq:rpf}
    \left({\pfh}_t, \psi_i\right) + \left(\frac{1}{\tau} \pd{f}{\pfh}, \psi_i\right) + \left(\sigma^2 \nabla \pfh, \nabla\psi_i\right)=0,% -<\nabla\pfh\cdot\bn,\psi_i>=0,
\end{equation}
\begin{equation}\label{eq:rc}
    \left({\conch}_t, \psi_i\right) + \left(D \nabla\frac{\partial f}{\partial u_h}, \nabla \psi_i\right) =0,%-<D\nabla \pd{f}{\pfh}\cdot\bn,\psi_i>=0,
\end{equation}
and
\begin{equation}\label{eq:rtemp}
\left({\temph}_t,\psi_i\right)+ \left(\alpha\nabla \temph,\nabla\psi_i\right)-\left(G_1,\psi_i\right)=0,%-<\nabla\temph\cdot\bn,\psi_i>=0,
\end{equation}
for all $\psi_i\subset V_h$, where $(\pf,\psi)=\int_{\Of}\pf\cdot\psi\dif{\Of}$ depicts the inner product in $L^2(\Of)$, and we have applied Green's theorem~\cite{braess01}. For the example problems in this manuscript, we approximate $V_h$ using bilinear quadrilateral (Q1) or biquadratic quadrilateral (Q2) element spaces in 2D and bilinear hexahedral element spaces in 3D~\cite{brenner2007fem,braess01}. 

\subsubsection{Time integration}\label{sec:time}
We discretize Eqs.~\eqref{eq:rpf}--\eqref{eq:rtemp} temporally with a weighted method and define the residual $\mathbf{F}=\left[F_1\ F_2\ F_3\right]^\mathrm{T}$ as follows:
\begin{multline}\label{eq:dpf}
 F_1\left(\pf_h^{n+1},\conc_h^{n+1},\temp_h^{n+1}\right)_i :=
    \frac{1}{\dt}\left(\pf_h^{n+1}-\pf_h^n,\psi_i\right) 
    + \thetam \, \left(\left(\frac{1}{\tau} \pd{f}{\pf}\right)^{n+1}, \psi_i\right) 
    + (1-\thetam)\,\left(\left(\frac{1}{\tau} \pd{f}{\pf}\right)^{n}, \psi_i\right)\\
    + \thetam \, \left(\sigma^2 \nabla \pfh^{n+1}, \nabla\psi_i\right) 
    + (1-\thetam)\, \left(\sigma^2 \nabla \pfh^{n}, \nabla\psi_i\right),
\end{multline}

\begin{multline}\label{eq:dc}
F_2\left(\pf_h^{n+1},\conc_h^{n+1},\temp_h^{n+1}\right)_i:=
    \frac{1}{\dt}\left(\conc_h^{n+1}-\conc_h^n,\psi_i\right)
    +\thetam\,\left(\left(D \nabla\pd{f}{\conc}\right)^{n+1}, \nabla \psi_i\right)\\
        +(1-\thetam)\,\left(\left(D \nabla\pd{f}{\conc}\right)^{n}, \nabla \psi_i\right),
\end{multline}
and
\begin{multline}\label{eq:dtemp}
F_3\left(\pf_h^{n+1},\conc_h^{n+1},\temp_h^{n+1}\right)_i:=
    \frac{1}{\dt}\left(\temp_h^{n+1}-\temp_h^n,\psi_i\right)
        +\thetam\,\left(\alpha\nabla \temp_h^{n+1},\nabla\psi_i\right)\\
            +(1-\thetam)\,\left(\alpha\nabla \temp_h^n,\nabla\psi_i\right)
    -\left(G_1(\widehat{\pf_h^n}),\psi_i\right),
\end{multline}
with
\begin{equation}\label{eq:theta}
\widehat{\pf_h^n}=\frac{\thetam}{\dt}\left(\pf_h^{n+1}-\pf_h^n\right)+\frac{1-\thetam}{\dt}\left(\pf_h^n-\pf_h^{n-1}\right).
\end{equation}
We use superscripts to denote time index, $\dt=t^{n+1}-t^n$, which is uniform timestep size from time $n$ to time $n+1$; $\thetam=0$ corresponds to the explicit (forward) Euler method, $\thetam=1$ corresponds to the implicit (backward) Euler method, and $\thetam=1/2$ corresponds to the implicit trapezoidal method (often referred to as Crank-Nicolson)~\cite{lambert:1991}. The subscript $i$ refers to vector index.
To determine $\pf_i^{n+1}$, $\conc_i^{n+1}$ and $\temp_i^{n+1}$,
we solve a fully-coupled, nonlinear system of
\begin{equation}\label{eq:nls}
\mathbf{F}\left(\bU\right) = 0,
\end{equation}
with
\begin{equation}
\mathbf{F}(\bU)=\left[F_1\ F_2\ F_3\right]^\mathrm{T}
\end{equation}
and
\begin{equation}
\bU_i=\left[\pf_i^{n+1}\ \conc_i^{n+1}\ \temp_i^{n+1}\right]^\mathrm{T},
\end{equation}
at each timestep. In this manuscript, we have chosen a $\thetam$-method with fixed timestep size to properly assess and analyze the algorithmic performance of time integration, linear and nonlinear solvers, and our preconditioning strategy. Our approach is to approximate a solution to Eq.~\eqref{eq:nls} with a Jacobian-free Newton-Krylov method discussed in Sec.~\ref{sec:jfnk}.

\subsection{Jacobian-free Newton-Krylov (JFNK) methods}\label{sec:jfnk}

We briefly describe the JFNK method. More detailed analyses can be found in~\cite{knoll:2004,kelley95}. The JFNK method for the discrete problem is
\begin{equation}\label{eq:residual}
  \mathbf{F}(\bU) = 
  \mathbf{0}, 
\end{equation}
with $\mathbf{F}:\mathbb{R}^N\to\mathbb{R}^N$, where $N$ is the number of unknowns. Given the Jacobian, $\Jac(\bU)$, with $\Jac(\bU)_{ij}=\partial\mathbf{F}(\bU)_i/\partial\bU_j$, it is straightforward to express the Newton iteration,
\begin{equation}
  \Jac(\bU^{(l)}) \, \delta \bU^{(l)} = -\mathbf{F}(\bU^{(l)})
\label{eq:jacobian-linear-system}
\end{equation}
and
\begin{equation}
  \bU^{(l+1)} \longleftarrow \; \bU^{(l)} + \Lambda \,\delta \bU^{(l)},
  \label{eq:newton-update}
\end{equation}
where the superscript $(l)$ denotes the iteration count of the Newton iteration, and $\delta \bU^{(l)}=\bU^{(l+1)}-\bU^{(l)}$ is the Newton step. The step length, $\Lambda\in(0,1]$, is chosen by a backtracking line search strategy, such that
\begin{equation}\label{eq:glob}
\|\mathbf{F}(\bU^{(l)}+\Lambda \,\delta \bU^{(l)})\|_2\le\|\mathbf{F}(\bU^{(l)})\|_2.
\end{equation}
Typically, $\Lambda$ is initially chosen as $\Lambda=1$ and is reduced until Eq.~\eqref{eq:glob} is satisfied. This process is also known as globalization~\cite{dennis:1996,pawlowski:2006}. The Newton iteration is terminated when
\begin{equation}\label{eq:tolerance}
\|\mathbf{F}(\bU^{(l)})\|_2<\tau_r\|\mathbf{F}(\bU^{(0)})\|_2+\tau_a,
\end{equation}
where $\tau_r$ is a specified relative tolerance, $\tau_a$ is a specified absolute tolerance, and $\|\cdot\|_2$ is the Euclidean norm on $l^2$.
 
Newton's method is impractical for solving large systems due to the high computational cost associated with forming the Jacobian $\Jac(\bU)$. Therefore, Krylov iterative solvers such as generalized minimum residual (GMRES) algorithm~\cite{saad95} are used to solve the linear system in Eq.~\eqref{eq:jacobian-linear-system}. This approach does not require the Jacobian matrix itself, but utilizes the operation of the Jacobian matrix on a vector $\mathbf{v}$. Approximating this matrix-vector product by a finite difference is the basis of the JFNK method.  
Specifically, to evaluate the matrix-vector product $\Jac(\bU^{(l)})\mathbf{v}$, a finite-difference approach, 
\begin{equation}
   \Jac(\bU^{(l)})\mathbf{v} 
      \approx \frac{1}{\jfeps} \left(\mathbf{F}(\bU^{(l)} + \jfeps\mathbf{v}) -\mathbf{F}(\bU^{(l)})\right)
                     ,
    \label{eq:JFmatvec}
\end{equation}
is commonly used~\cite{knoll:2004,pernice:1998}, where $\jfeps$ is chosen heuristically to avoid problems with machine precision~\cite{brown90}. As $\mathbf{F}(\bU^{(l)})$ is readily available in the JFNK method, each GMRES iteration only requires an evaluation of $\mathbf{F}(\bU^{(l)} + \jfeps\mathbf{v})$.

An inexact convergence tolerance is used on the linear solve for the JFNK method~\cite{pernice:1998,Dembo1982,walkerIN}. The tolerance to which we solve the linear problem on each nonlinear iteration is related to the current nonlinear residual, $\mathbf{F}(\bU^{(l)})$, by
\begin{equation}\label{eq:force1}
\| \Jac(\bU^{(l)}) \, \delta \bU^{(l)} +\mathbf{F}(\bU^{(l)})\|_2\le\eta_{(l)}\|\mathbf{F}(\bU^{(l)})\|_2,
\end{equation}
with
\begin{equation}
\eta_{(l)} = \gamma\left(\frac{\|\mathbf{F}(\bU^{(l)})\|_2}{\|\mathbf{F}(\bU^{(l-1)})\|_2}\right)^\omega
\end{equation}
and 
\begin{equation}\label{eq:forcing}
\max(\gamma\eta_{(l-1)}^{\omega},\eta_{\min})\le\eta_{(l)}\le\eta_{\max}.
\end{equation}
Equations~\eqref{eq:force1}--\eqref{eq:forcing} are commonly referred to as the forcing term, and its particular choice is critical to efficiency of the JFNK method. 
Our choice of values for $\gamma$, $\omega$, $\eta_{(0)}$, $\eta_{\min}$ and $\eta_{\max}$ are provided in Sec.~\ref{sec:metal}.

\subsection{Preconditioning}\label{sec:precon}
The dominant cost of the JFNK method is GMRES iteration. To reduce the number of GMRES iterations, preconditioning is used. We apply physics-based preconditioning~\cite{chacon:2002a,knoll:2000a} as a right preconditioning process $\Prec^{-1}$, and the matrix-vector product expression in Eq.~\eqref{eq:JFmatvec} becomes
\begin{equation}
   \Jac(\bU^{(l)})\Prec^{-1}\mathbf{v} 
      \approx \frac{1}{\jfeps}\left( \mathbf{F}(\bU^{(l)} + \jfeps \Prec^{-1}\mathbf{v}) -\mathbf{F}(\bU^{(l)})\right).
    \label{eq:preJFmatvec}
\end{equation}
The operator $\Prec^{-1}$ is chosen such that $\Prec^{-1}\mathbf{v}$ is a suitable approximation to $\Jac(\bU^{(l)})^{-1}\mathbf{v}$.  In addition, $\Prec^{-1}\mathbf{v}$ should be inexpensive to construct and apply.

We introduce a process for applying the preconditioner. As Eqs.~\eqref{eq:rpf}--\eqref{eq:rtemp} constitute a parabolic system of PDEs, we address dominant time scales that are important for constructing our preconditioner and examine a linearized form of the Jacobian to arrive at the preconditioning operator $\bM^{-1}$.

\subsubsection{Time scales}\label{sec:timescale}
The time scales associated with diffusion in Eqs.~\eqref{eq:rpf}--\eqref{eq:rtemp} are given by:
  \begin{equation}\label{eq:dtpf}
    \Dt_\pf \leq  C_\pf\,\tau\, h^2 / W^2,
  \end{equation}
  \begin{equation}\label{eq:dtc}
    \Dt_\conc \leq  C_\conc\, h^2 / D,
  \end{equation}
  and
\begin{equation}\label{eq:dtt}
    \Dt_{\temp} \leq C_{\temp}\, h^2 / \alpha,
  \end{equation}
 respectively;
where $h$ is associated with mesh size and $C_\pf$, $C_\conc$, and $C_\temp$ are constants independent of $h$. Equations~\eqref{eq:dtpf}--\eqref{eq:dtt} determine the maximum stable timestep for explicit methods~\cite{lambert:1991} applied to Eqs.~\eqref{eq:dpf}--\eqref{eq:dtemp}. 
We additionally note that timestep size, $\Delta t$, and mesh size, $h$, are explicitly coupled and related to the condition number of the linearized and discretized Jacobian through Eqs.~\eqref{eq:dtpf}--\eqref{eq:dtt}~\cite{lambert:1991}.
In typical material processes, $\Dt_\pf > \Dt_\conc > \Dt_T$ is valid~\cite{provatas2011}.
 
The separation between these time scales presents challenges for time integration. We call a time integration method \emph{algorithmically scalable} if the CPU time is inversely proportional to the timestep size. Explicit time integration to a fixed computation time is inherently algorithmically scalable; if the timestep size is doubled, CPU time is reduced by a half. A drawback of explicit methods is the stability restriction of the maximum timestep size. Explicit methods for phase-field simulations are unstable for timesteps larger than $\Dt_T$. While implicit methods have superior stability properties, when applied to problems with nonlinear diffusion, conditioning of the linear system typically suffers when $\dt>\Dt_\temp$, requiring many linear iterations for convergence of the linear solver. This is particularly true for Eqs.~\eqref{eq:dpf}--\eqref{eq:dtemp}. Hence, implicit time integration methods, in general, do not scale well algorithmically unless effectively preconditioned.

Our goal is to integrate the fully-coupled system of Eqs.~\eqref{eq:dpf}--\eqref{eq:dtemp} with an implicit time integration with timesteps $\Dt>\Dt_\pf$ (Eq.~\eqref{eq:dtpf}) that are larger than the explicit timestep limit $\Dt_\temp$, while maintaining the desired level of accuracy and algorithmic scalability.

\subsubsection{Construction of preconditioner}
We construct a preconditioning operator, $\bM^{-1}$, that is effective when $\Dt>\Dt_\temp$. Recall that $\bM^{-1}$ has the properties that $\bM^{-1}\bv$ is a suitable approximation to $\Jac(\bU^{(l)})^{-1}\mathbf{v}$, and $\bM^{-1}\bv$ should be inexpensive to construct and apply. We drop superscripts and consider the Jacobian of Eqs.~\eqref{eq:dpf}--\eqref{eq:dtemp}, 
\begin{equation}\label{eq:jac}
\Jac(\bU)
 = 
\begin{bmatrix} \frac{\partial F_1(\pf, \conc, \temp)}{\partial\pf} & \frac{\partial F_1(\pf, \conc, \temp)}{\partial \conc} & \frac{\partial F_1(\pf, \conc, \temp)}{\partial\temp} \\
\frac{\partial F_2(\pf, \conc, \temp)}{\partial\pf} & \frac{\partial F_2(\pf, \conc, \temp)}{\partial \conc} & \frac{\partial F_2(\pf, \conc, \temp)}{\partial\temp} \\
\frac{\partial F_3(\pf, \conc, \temp)}{\partial\pf} & \frac{\partial F_3(\pf, \conc, \temp)}{\partial \conc} & \frac{\partial F_3(\pf, \conc, \temp)}{\partial\temp}
\end{bmatrix}:=
\begin{bmatrix}
M_{11} & M_{12} & M_{13} \\
M_{21} & M_{22} & M_{23} \\
M_{31} & M_{32} & M_{33} \\
\end{bmatrix},
\end{equation}
and construct $\bM$ as an approximation to $\Jac(\bU)$ by discarding off-diagonal blocks to arrive at 
\begin{equation}\label{eq:pre}
\bM=\mbox{diag}(M_{11},M_{22}, M_{33}).
\end{equation}
By ignoring off-diagonal terms, we have accepted that our preconditioner will be less effective when the solver uses timesteps approaching small time scales. Note that inversion of Eq.~\eqref{eq:pre} is equivalent to one iteration of the block-Jacobi method~\cite{kelley95}. In our examples, the matrix in Eq.~\eqref{eq:pre} is approximately inverted blockwise using algebraic multigrid methods~\cite{briggs2000multigrid,hackbusch2013multi,trottenberg.ea00}. In particular, we utilize the \texttt{ML}~\cite{gee:2006} and \texttt{MueLu}~\cite{MueLu,MueLuURL} packages within the \texttt{Trilinos} library~\cite{heroux:2005,trilinos}.

%-------------------------------------------------------------

\section{Tusas}\label{sec:features}

To mitigate the challenges due to time scales, length scales, and their coupling, we have designed \tusas~\cite{tusas} with the following capabilities:

\begin{itemize}
\item \textit{Higher-order discretizations:} \tusas\ utilizes an unstructured finite element spatial discretization, which can be more flexible and for arbitrary mesh geometries (such as curved boundaries meltpool boundaries found in AM solidification simulations) compared to finite difference or finite volume methods~\cite{brenner2007fem,braess01}. Discretizations consist of linear and bilinear quadrilateral, triangular, tetrahedral, or hexahedral Lagrange (nodal) elements in 2D and 3D. Meshing is facilitated through the \texttt{Exodus II} library~\cite{exodusii07,exodus:96}. Time integration techniques implemented in \tusas\ include explicit and implicit Euler, Crank-Nicolson, general second-order backward difference formula, and methods based on implicit midpoint rule. Implicit methods possess less restrictive timestep stability requirements, allowing for larger timesteps than explicit methods; and, as we show in Sec.~\ref{sec:alg}, can provide less computational cost for a given level of accuracy than explicit methods.

\item \textit{Robust algorithms and solvers:} \tusas\ leverages the \texttt{Trilinos} library~\cite{trilinos} and utilizes JFNK nonlinear solvers provided by \texttt{NOX}~\cite{kolda:nox} and GMRES linear solvers provided by \texttt{Belos}~\cite{bavier:2012} and \texttt{AztecOO}~\cite{heroux:2004}. The key to an efficient implementation of JFNK is effective preconditioning. As the dominant cost of JFNK is the linear solver, Eq.~\eqref{eq:jacobian-linear-system}, effective preconditioning reduces the number of linear solver iterations per Newton iteration and allows larger timesteps by \emph{stepping over} stiff physics common to elliptic operators. Thus, the timestep is determined by accuracy instead of by numerical stability constraints~\cite{knoll:2004,knoll:2005,park:2010}. Our preconditioning strategy is based on block factorization and algebraic multigrid (AMG) methods using the \texttt{ML}~\cite{gee:2006} and \texttt{MueLu}~\cite{MueLu,MueLuURL} libraries. Although not addressed in this manuscript, \tusas\ additionally implements an interface to the \texttt{Rythmos}~\cite{rythmos} adaptive time integration library in \texttt{Trilinos}. We show specifically in Sec.~\ref{sec:alg} and Sec.~\ref{sec:performance} that the use of these packages and algorithms allow for efficient implicit time integration while affording algorithmic scalability.

\item \textit{Modular object-oriented programming:} The combination of finite element discretization and JFNK allows for a unique opportunity to leverage modular, object-oriented software development. The JFNK method only requires evaluation of the residual, Eqs.~\eqref{eq:rpf}--\eqref{eq:rtemp}, realized on each element \textit{via} pointwise evaluation at Gauss points. Hence, from an implementation standpoint, \tusas\ allows a flexible framework as it only requires the user to implement code for each block of the residual equation, independent of dimension, and supports a variety of easily coupled phase-field models. Optionally, \tusas\ allows the user to implement code for each block row of the preconditioning matrix, Eq.~\eqref{eq:pre}. \tusas\ is written in C++ and provides a flexible interface that enables runtime configuration and coupling of a large set of arbitrary PDEs and complex multiphysics facilitated by pointers to functions precompiled on the device, an approach similar to~\cite{gaston:2009a}. Since it is impractical to implement new code every time a set of similar or additional physics is required to solve, the \tusas\ framework leverages effective code reusability, allowing rapid development of the code without requiring extensive modification. 
In addition to \texttt{Trilinos}, \tusas\ makes heavy utilization of \texttt{CMake}~\cite{cmake}, \texttt{Boost}~\cite{boost}, \texttt{HDF5}~\cite{hdf5}, and \texttt{NetCDF}~\cite{rew:1990}.

\item \textit{Massively parallel computing:} \tusas\ is specifically designed to leverage hierarchical parallelism on emerging heterogeneous hardware and architectures. In particular, \tusas\ utilizes the \texttt{Kokkos} library~\cite{kokkos,kokkos_paper}. \texttt{Kokkos} provides an abstract interface for shared memory to both the \texttt{OpenMP} library~\cite{Openmp} for threading on CPU hardware, and the \texttt{CUDA} library~\cite{Cuda} for execution on GPU hardware. We have specifically chosen the \texttt{Kokkos} and \texttt{Trilinos} software stack due to its inherent exascale readiness and scalability on GPUs that is not readily mature in other solver distributions such as~\cite{gaston:2009a,falgout02,hindmarsh2005sundials}. Execution on multi-node, distributed memory hardware for multi-CPU and multi-GPU is afforded by \texttt{MPI}~\cite{Mpich}. 
%As we are designing high-fidelity microstructure simulations, we are observing an increasing need for exascale computing (one quintillion ($10^{18}$) floating-point operations per second).
By utilizing \texttt{MPI} and \texttt{Kokkos}, \tusas\ is uniquely targeting emerging exascale architectures. We demonstrate the ability of \tusas\ to scale both strongly and weakly with up to 4 billion unknowns on thousands of GPUs in Sec.~\ref{sec:performance}.

\end{itemize}

%We refer the reader to the \texttt{GitHub} page (\url{https://github.com/chrisknewman/tusas})~\cite{tusas} for instructions on how to obtain, install, compile, and execute \tusas. 
%---------------------------------------------------------------------------------------

\section{Results}\label{sec:results}
To demonstrate the performance and flexibility of \tusas\ framework, we examine two examples from phase-field simulation of solidification. The first example is free dendrite growth in a homogeneous pure metal melt. In this example, we focus on algorithmic performance, specifically in terms of accuracy and efficiency. The second example is columnar dendrite growth during directional solidification of a dilute binary metal alloy; here, we examine parallel scalability and efficiency of 3D simulations on emerging supercomputer architectures.

\subsection{Free growth}\label{sec:metal}

Dendritic solidification in a homogeneous material (in the absence of concentration) is well-understood community-determined reference benchmark problem to validate phase-field frameworks~\cite{jokisaari2018,stephen2020}. To model this test problem, we adopt the mathematical forms of the nonlinear equations for $\pf$ and $T$ given in~\cite{cummins:2002}. Applying non-dimensionalization with a particular choice of $f(\pf, \temp)$, Eq.~\eqref{eq:dpf} becomes
\begin{multline}\label{eq:pf1}
F_1\left(\pf_h^{n+1},\temp_h^{n+1}\right)_i=
    \frac{1}{\dt}\left(\gs^2(\atheta^{n+1})\pf_h^{n+1}-\gs^2(\atheta^n)\pf_h^n,\psi_i\right)\\
    +\thetam\,\beta\,\Gamma\, \left(\gs^2(\atheta^{n+1})\nabla\pf_h^{n+1},\nabla\psi_i\right)
   +(1-\thetam)\,\beta\,\Gamma\, \left(\gs^2(\atheta^n)\nabla\pf_h^n,\nabla\psi_i\right)\\
    +\frac{\thetam}{2}\left(\|\nabla\pf_h^{n+1}\|\nabla_{\nabla\pfh}\gs^2(\atheta^{n+1}),\nabla\psi_i\right)
   +\frac{1-\thetam}{2}\left(\|\nabla\pf_h^n\|\nabla_{\nabla\pfh}\gs^2(\atheta^n),\nabla\psi_i\right)\\
    +\thetam\,\left(G_2(\pf_h^{n+1},\temp_h^{n+1}),\psi_i\right)
    +(1-\thetam)\,\left(G_1(\pf_h^n,\temp_h^n),\psi_i\right),
\end{multline}
with
\begin{equation}\label{eq:g1}
G_2(\pfh,\temph)=\beta\,\Gamma\,\frac{\pfh(1-\pfh)(1-2\pfh)}{h^2}
-5\,\beta\,(\temp_m-\temph)\frac{\pfh^2(1-\pfh)^2}{h},
\end{equation}
and the operator
\begin{equation}
  \nabla_{\nabla \pf}= \left[\frac{\partial}{\partial\pf_x}\ \frac{\partial}{\partial\pf_y}\ \frac{\partial}{\partial\pf_z}\right]^{\mathrm{T}}.
  \end{equation}
Here,
\begin{equation}\label{eq:gs}
   \sigma=\gs(\atheta)=1-3\epsilon+4\epsilon\frac{\pf_x^4+\pf_y^4+\pf_z^4}{\|\pf\|^4} 
\end{equation}
is the four-fold solid-liquid interface anisotropy with magnitude $\epsilon$. Similarly, the dimensionless form of Eq.~\eqref{eq:dtemp} for $T$ becomes
\begin{equation}\label{eq:temp1}
F_3\left(\pf_h^{n+1},\temp_h^{n+1}\right)_i=
    \frac{1}{\dt}\left(\temp_h^{n+1}-\temp_h^n,\psi_i\right)\\
    +\thetam\,\left(\alpha\nabla \temp_h^{n+1},\nabla\psi_i\right)
    +(1-\thetam)\left(\alpha\nabla \temp_h^n,\nabla\psi_i\right)
    -\left(G_1(\widehat{\pf_h^n}),\psi_i\right).
\end{equation}
We utilize 9 quadrature points in each quadrilateral element to accurately integrate the nonlinear terms in Eqs.~\eqref{eq:pf1}--\eqref{eq:temp1}. Initial conditions consist of a perturbed circular seed given by:
\begin{equation}\label{eq:circle}
\pf(\bx,0)= \left\{
    \begin{array}{rcc} 
    1&:& \|\bx\|\le r\\
    0&:&  \|\bx\|> r
    \end{array} 
    \right.
    ,
\end{equation}
with radius $r = 0.3\,\gs(\atheta)$ and $\temp(\bx,0)= \temp_m$. We recall that $\pf = 1$ corresponds to solid and $\pf = 0$ corresponds to liquid. The dimensionless material parameters utilized for simulations are listed in Table~\ref{tab:parameters}. In this model, the length scale is $d_0$ and the time scale is $\tau = d_0^2/\alpha$.
\begin{table}[!t]
  \begin{centering}\footnotesize
    \begin{tabular}{ccccccccc}\hline
      $\epsilon$&Anisotropy& $0.005$&&&&
      %$M$&lobe number& $4.0$\\
      $\Omega$&Domain& $[0,4.5]\times[0,4.5]$ \\
      $\alpha$&Thermal diffusivity& $4.0$ &&&&
      $L/c_p$&Latent heat&$1.0$\\
      $\beta$&Kinetic coefficient& $1/\Gamma=191.82$&&&&
      $\temp_\infty$&Liquidus temperature& $1.0$\\
      $W$&Interface width&$h/0.4=0.02652$&&&&
      $\Delta$ & Undercooling & $(\temp_m-\temp_\infty)/(L/c_p)=.55$ \\
      $d_0$&Capillary length&$.139\,W$&&&&
\\ \hline
    \end{tabular}
    \caption{Nondimensional material properties and simulation parameters in Eqs.~\eqref{eq:pf1}--\eqref{eq:temp1}, adopted from~\cite{cummins:2002}.}
    \label{tab:parameters}
  \end{centering}
\end{table}

Simulations were performed on $\Of = [0,\ 4.5]^2$ consisting of a structured mesh with $300 \times 300$ quadrilateral elements with mesh spacing $h = 0.03$. To study convergence and algorithmic performance, we compare explicit-Euler ($\thetam = 0$) and implicit Crank-Nicolson ($\thetam = 1/2$) methods, with $\Delta t$ varied between $10^{-3}$ and $10^{-6}$. In addition, we compare the effects of bilinear (Q1) and biquadratic (Q2) discretizations on anisotropy. We utilize the following nonlinear solver parameters: $\eta_{0} = 0.1$, $\eta_{\min} = 10 ^{-6}$, $\eta_{\max} = 0.01$, $\omega = 1.5$, $\gamma = 0.9$, $\tau_a = 0$, and $\tau_r = 10^{-6}$ for all simulations (Sec.~\ref{sec:jfnk}).

\begin{figure}[t]
\centering
\subfloat[]{\includegraphics[scale=0.085]{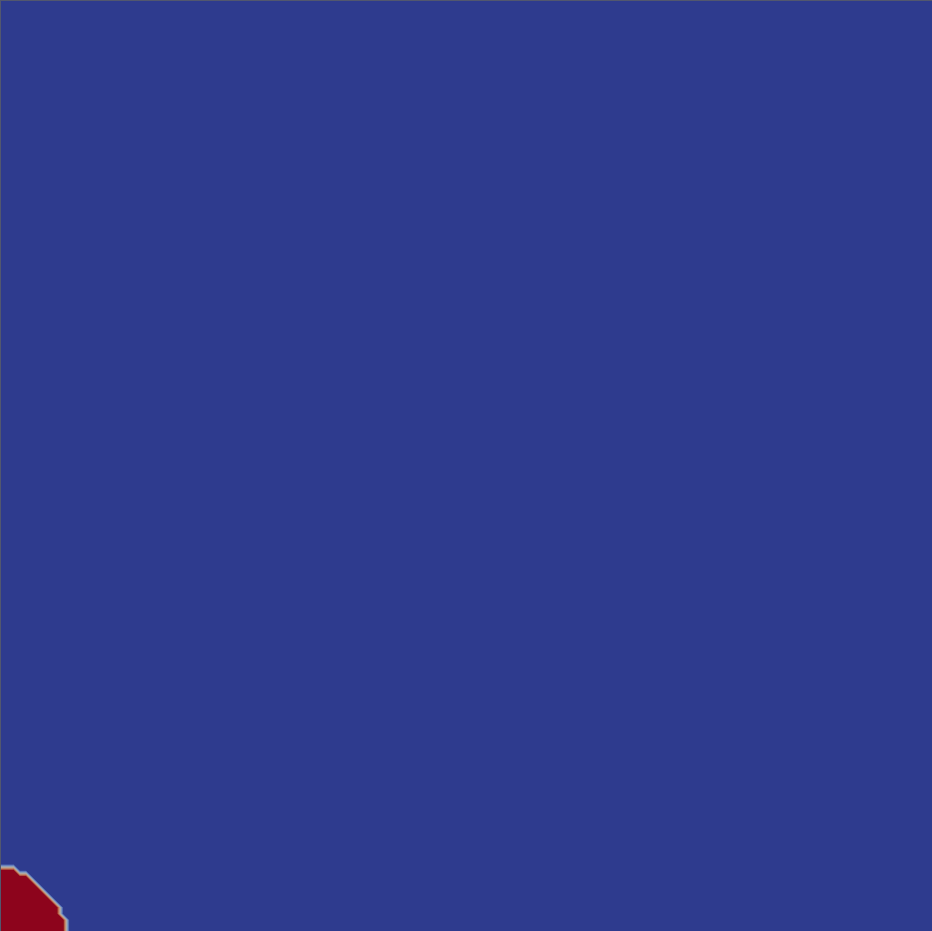}}\hfill
\subfloat[]{\includegraphics[scale=0.072]{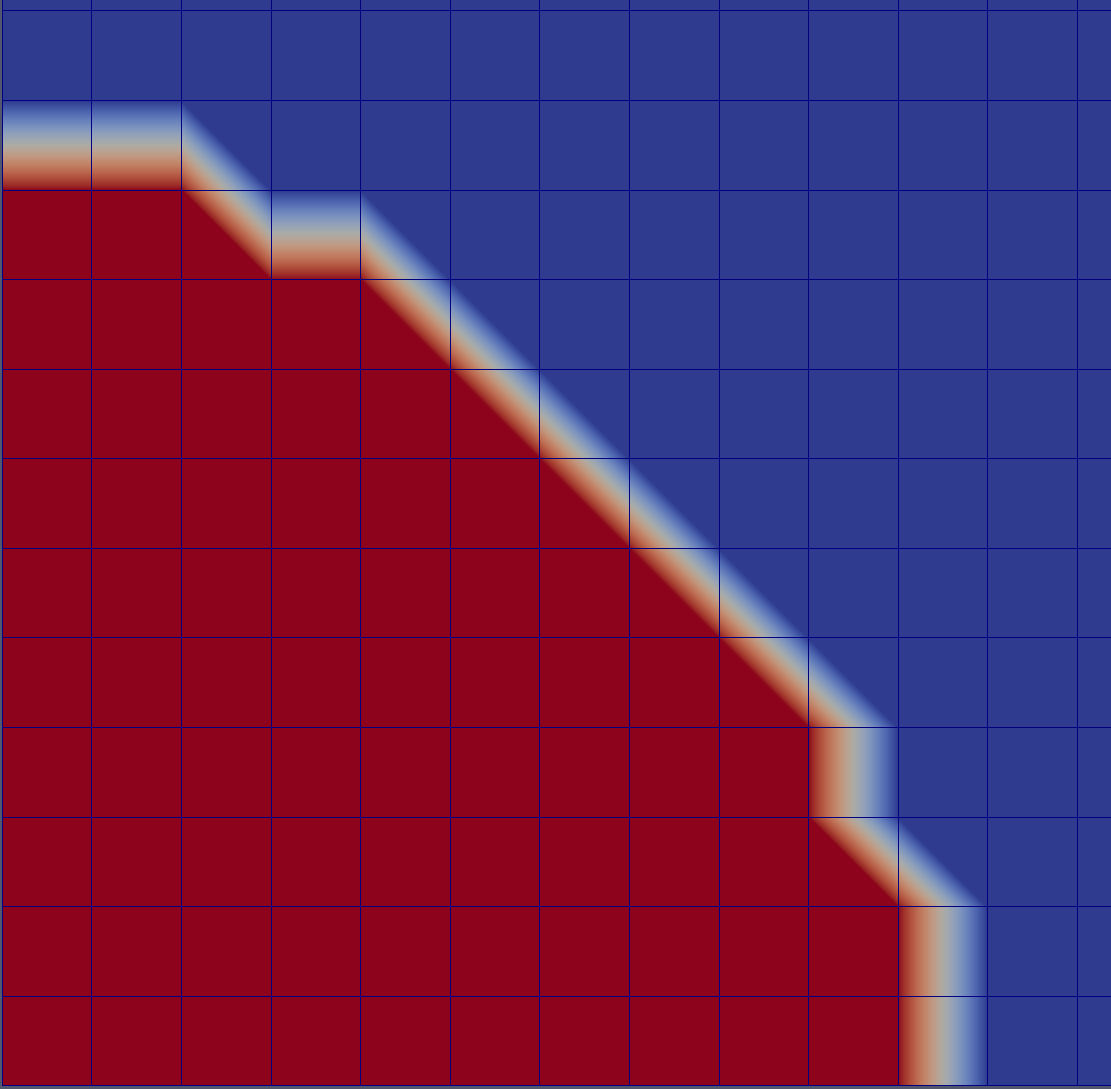}}\hfill
\subfloat[]{\includegraphics[scale=0.6]{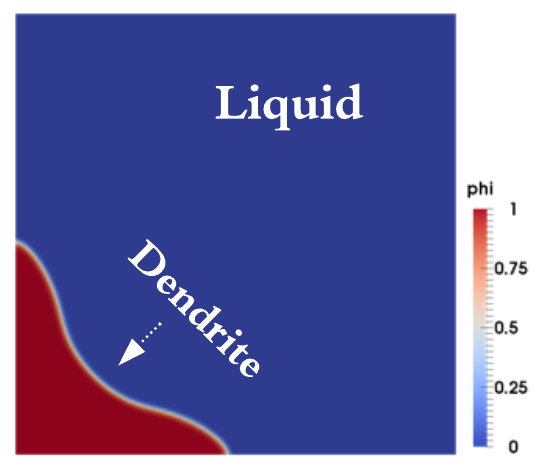}}\hfill
\subfloat[]{\includegraphics[scale=0.31]{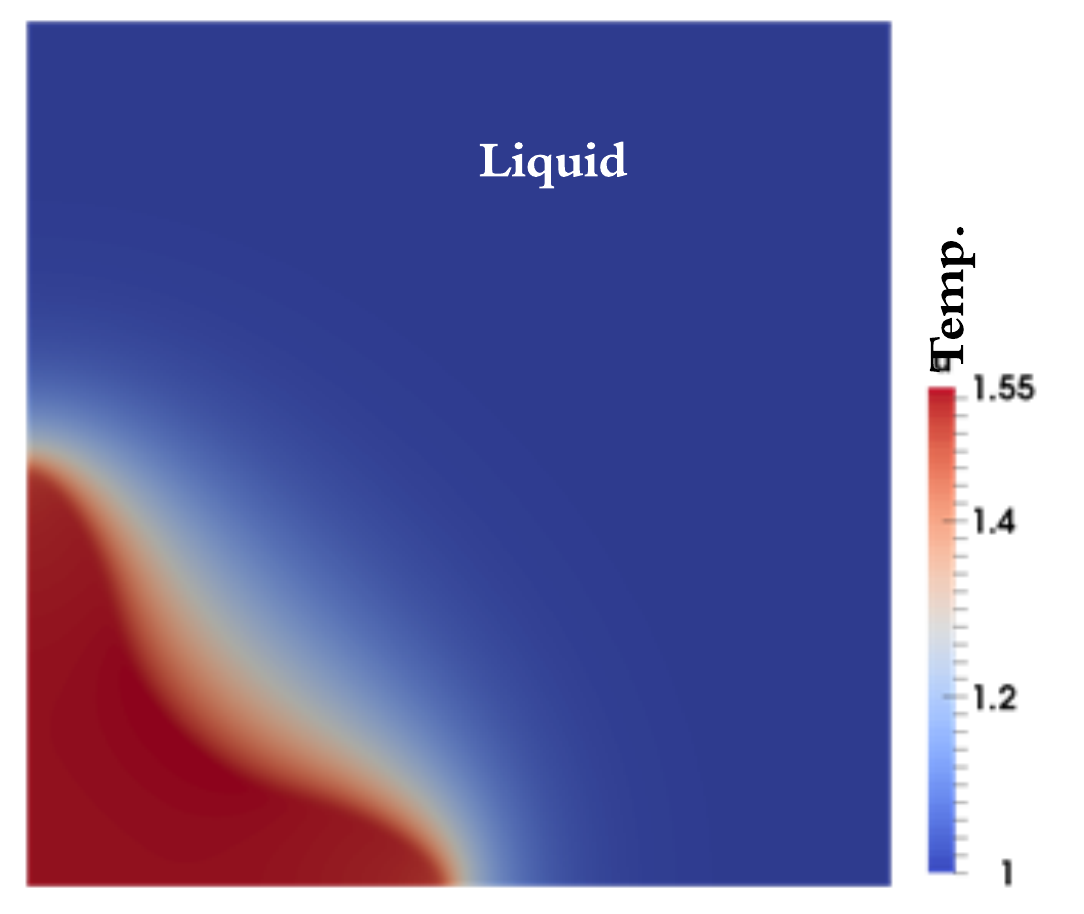}}
\caption{(a) Initial condition. (b) An enlarged view of the circular seed in a background finite element mesh. (c) Visualization of the final phase-field solution $\pf$ at the steady-state. (d) Visualization of the temperature field $T$ at the steady-state.}
\label{fig_dendrite}
\end{figure}

Simulations were carried out to dimensionless time $t_f$ = 0.14, far beyond dendrite tip reaching steady-state velocity, thus allowing a direct comparison with the analytical sharp-interface steady-state solutions for verification. Note that the number of timesteps required to attain $t_f$ = 0.14 would be different for different temporal approaches used in our simulations, since different values of $\Delta t$ are required for these approaches due to CFL constraints (Sec.~\ref{sec:timescale}). The initial solid grows to the typical dendrite shape in Fig.~\ref{fig_dendrite}. In particular, the initial condition, $\pf(\bx, 0)$, is shown in Fig.~\ref{fig_dendrite}a and is enlarged in Fig.~\ref{fig_dendrite}b to illustrate the background finite element mesh. Figure~\ref{fig_dendrite}c shows the steady-state solution of $\pf$, and steady-state $\temp$ is shown in Fig.~\ref{fig_dendrite}d. 

\subsubsection{Verification and validation}

It is essential to perform convergence tests of the \tusas\ solutions to verify the temporal and spatial discretizations and to further verify with the available physics-based model analytical predictions. To demonstrate temporal convergence, we define error for $\bU_i = [\pf_i\ \temp_i]^\mathrm{T}$ as $\|\bU -\bU_{\text{ref}}\|_{2}$, where $\bU_{\text{ref}}$ is a reference solution computed with the Crank-Nicolson method with ${\Delta t}_{\text{ref}} = 1.953125\times 10^{-6}$. This value of ${\Delta t}_{\text{ref}}$ was chosen such that a sufficiently high-resolution reference solution could be used for a verification of second-order convergence for the Crank-Nicolson method and first-order convergence of the explicit Euler method. 

\begin{figure}[t]
\centering
\includegraphics[scale=0.6]{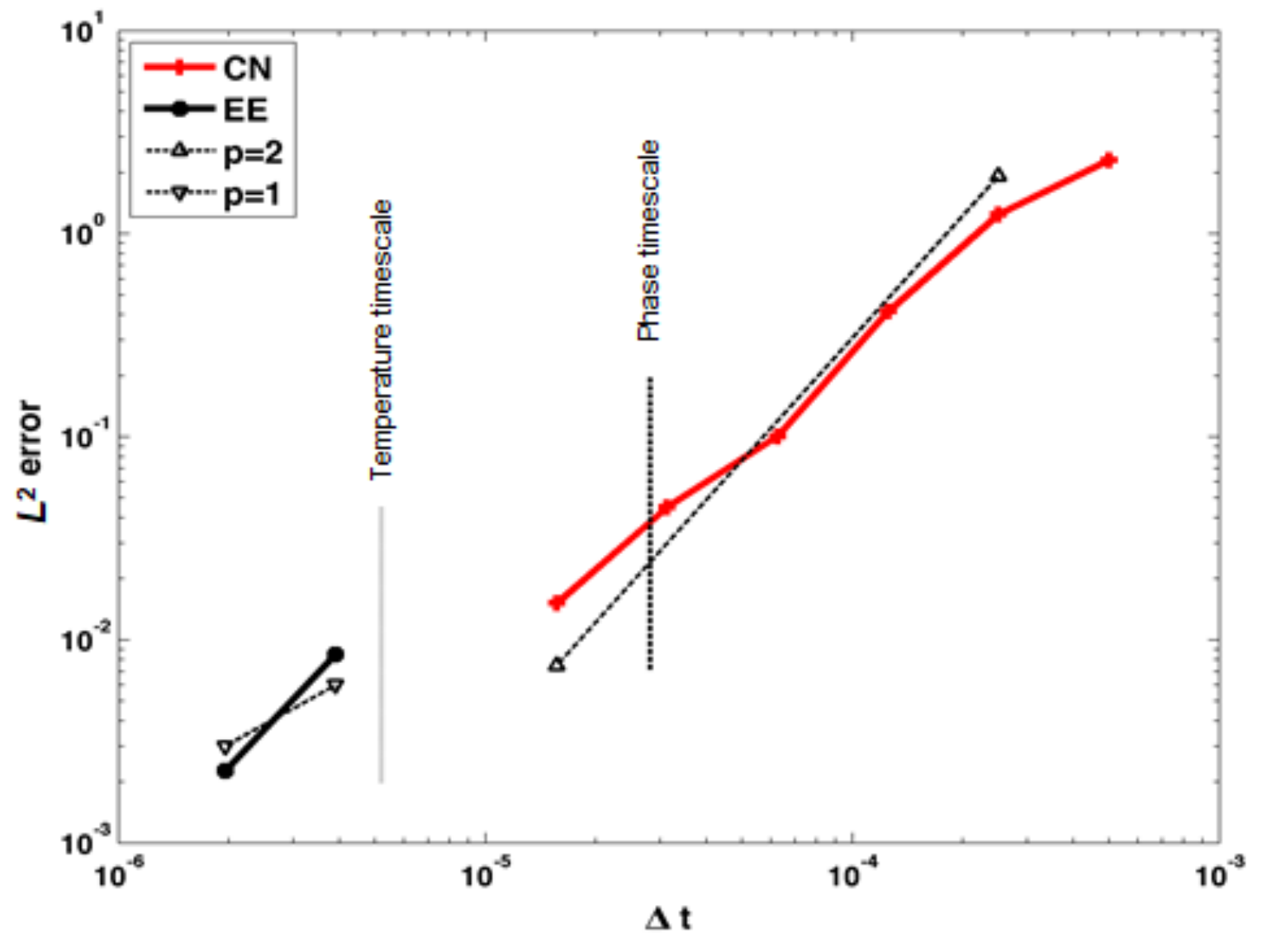}
\caption{Verification of temporal convergence rates for Crank-Nicolson (CN) and explicit Euler (EE). Reference lines for second-order ($p = 2$) and first-order ($p = 1$) convergence and reference lines for $\dt=\Dt_\temp$ (Temperature timescale) and $\dt=\Dt_\pf$ (Phase timescale) are shown.}
\label{fig_error}
\end{figure}

Figure~\ref{fig_error} shows error as a function of the timestep size at $t = t_f$ for explicit-Euler (EE) and implicit Crank-Nicolson (CN) methods, with $\Of$ spatially discretized with $22,500$ bilinear Lagrange (Q1) finite elements. Figure~\ref{fig_error} also provides a verification of order-of-accuracy for each method. In particular, the EE method is first-order accurate with timestep size restricted by the CFL condition associated with thermal diffusion $\dt<\Dt_\temp$ (Eq.~\eqref{eq:dtt}). In contrast, the implicit CN method is second-order accurate, and allows timesteps much larger than the CFL condition associated with phase diffusion $\dt>\Dt_\pf$ (Eq.~\eqref{eq:dtpf}).

\begin{figure}[p]
\centering
\subfloat[]{\includegraphics[scale=0.7]{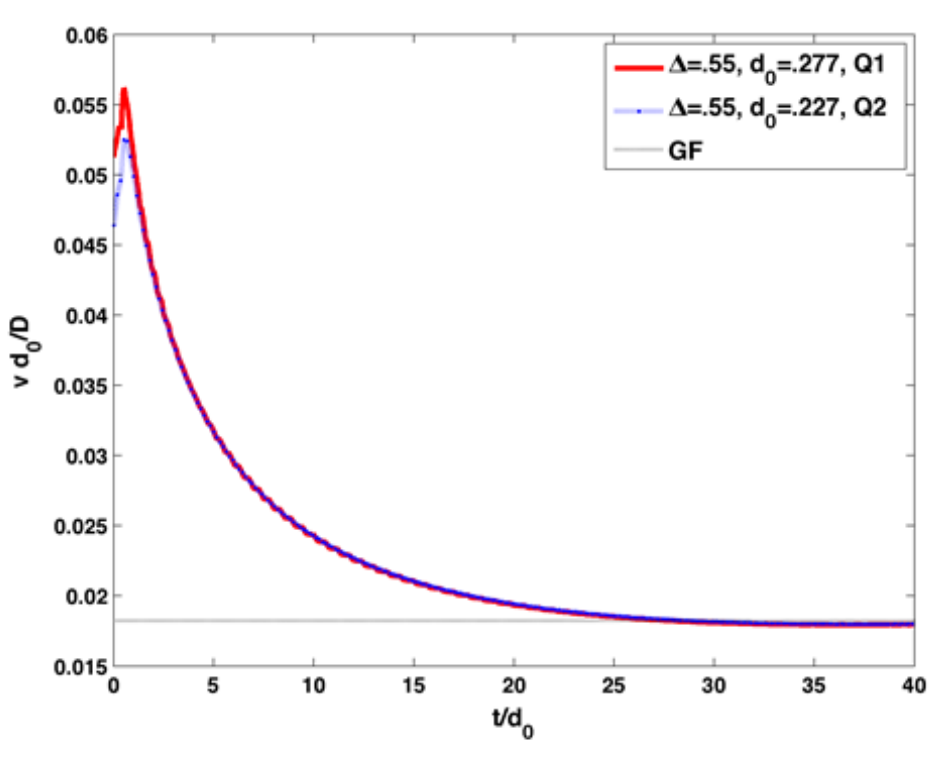}} \\
\subfloat[]{\includegraphics[scale=0.7]{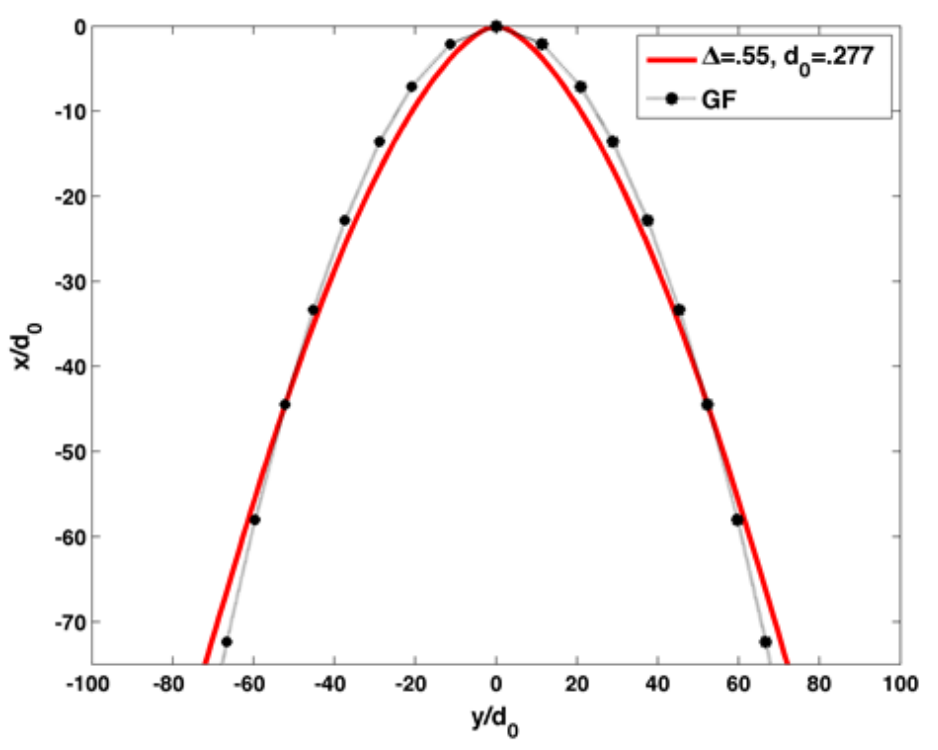}} \\
\subfloat[]{\includegraphics[scale=0.9]{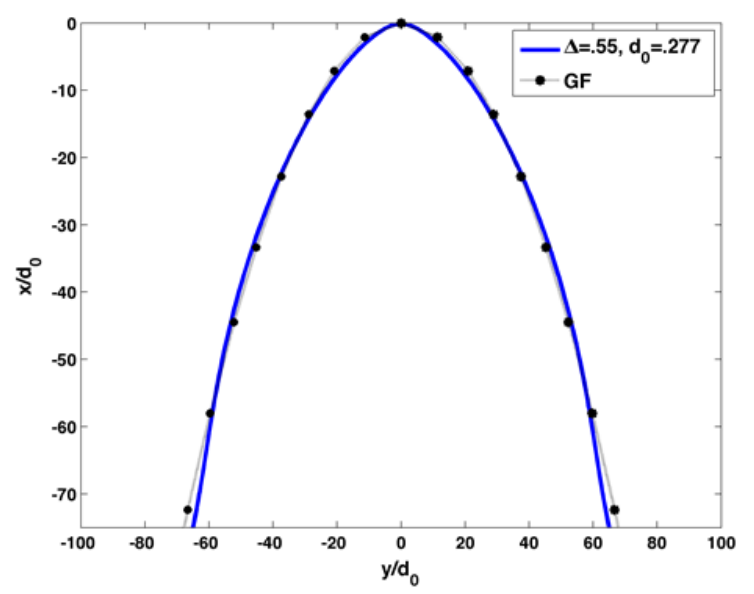}} 
\caption{(a) Evolution of dimensionless dendrite tip velocity, $\tilde{v}$, as a function of time for Q1 and Q2 elements with undercooling $\Delta = .55$ and capillary length $d_0 = .227$; evolution of dimensionless tip velocity $\tilde{v}_{\text{GF}}$ given by Green's function (GF). (b) Steady-state tip radius with Q1 element; tip radius given by GF. (c) Steady-state tip radius with Q2 element and GF.}
\label{fig_tip}
\end{figure}

To demonstrate spatial convergence, we make use of an analytic Green's function solution for our problem; this is valid at steady-state~\cite{kessler:1986} and provides a steady-state tip velocity as a function of undercooling $\Delta$ and capillary length $d_0$. We use this analytical solution to make a quantitative comparison with our Crank-Nicolson solution at $t = t_f$ and $\dt = \dt_{\text{ref}}$, utilizing both bilinear (Q1) and biquadratic (Q2) Lagrange spatial discretizations with $\Delta = .55$, $d_0 = .277$, and $22,500$ finite elements. In particular, Table~\ref{tab:tiperr} reports our simulated dimensionless steady-state tip velocity, $\tilde{v} = v\, d_0/\alpha$, the dimensionless steady-state tip velocity given by the Green's function solution, $\tilde{v}_{\text{GF}} = v_{\text{GF}}\, d_0/\alpha$, the absolute error, defined as $|\tilde{v}-\tilde{v}_{\text{GF}}|$, and relative error (\%), defined as $|\tilde{v}-\tilde{v}_{\text{GF}}|/\tilde{v}_{\text{GF}}\times 100$, for both Q1 and Q2 discretizations. The last column of Table~\ref{tab:tiperr} confirms the error decreases by a factor of more than 2 for Q2 compared to Q1. We note that the relative error in our calculations is similar or lower to that with other phase-field simulation frameworks for the same dendrite growth benchmark problem~\cite{stephen2020}. The evolution of tip velocity for both Q1 and Q2 is shown in Fig.~\ref{fig_tip}a, which confirms that $\tilde{v}$ reaches a fixed value upon the dendrite reaching the steady-state and shows qualitatively similar tip velocity evolution for both Q1 and Q2. 
Additionally, the Green's function solution provides a steady-state tip radius profile, which we use to qualitatively validate the Q1 and Q2 solutions. Figure~\ref{fig_tip}b shows dimensionless steady-state tip radius for the Q1 solution and the Green's function solution; Fig.~\ref{fig_tip}c shows the corresponding comparison for Q2. From a qualitative standpoint, the Q2 solution appears to more accurately resolve the Green's function solution and suggests that the higher-order biquadratic finite element may more accurately approximate the highly nonlinear anisotropic term $g_s(\atheta)$ (Eq.~\eqref{eq:gs}), critical for complex dendritic morphology evolution.

\begin{table}[!t]%\setlength{\tabcolsep}{30.0pt}
  \begin{centering}\footnotesize
    \begin{tabular}{ccccc}\hline
      Element & $\tilde{v}$ & $\tilde{v}_{\text{GF}}$  & $|\tilde{v}-\tilde{v}_{\text{GF}}|$ & Relative error \\ \hline
      Q1&.01822&.017&.00122&7\% \\
      Q2&.01754&.017&.00054&3\% \\ \hline
    \end{tabular}
    \caption{Dimensionless steady-state dendrite tip velocities,
$\tilde{v}$ and
$\tilde{v}_{\text{GF}}$, and error metrics for both Q1 and Q2 discretizations.}
    \label{tab:tiperr}
  \end{centering}
\end{table}

%--------------------------------------------------------------------
\subsubsection{Algorithmic performance}\label{sec:alg}

We precondition JFNK with the following forms of
\begin{equation}\label{eq:m11}
(M_{11})_{ij}=\frac{1}{\dt}(\gs^2(\atheta_j)\psi_j,\psi_i)+\thetam\,\beta\,\Gamma\, (\gs^2(\atheta_j)\nabla\psi_j,\nabla\psi_i)
\end{equation}
and 
\begin{equation}\label{eq:m22}
(M_{33})_{ij}=\frac{1}{\dt}(\psi_j,\psi_i)+
\thetam\,(\alpha\nabla \psi_j,\nabla\psi_i)
\end{equation}
in Eq.~\eqref{eq:pre} (with $(M_{22})_{ij}$ absent). For our examples, we utilize \texttt{Trilinos ML}~\cite{gee:2006} smoothed aggregation multigrid approach with 2 sweeps of symmetric Gauss-Siedel error smoother, 2 V-cycles, 4 levels of coarsening, and $45,602$ unknowns. Results were obtained in serial on MacBook Pro, OS X 10.9.4, 2.8 GHz Intel Core i7, 16 GB 1600 MHz DDR3, GCC 4.9.1. 

\begin{figure}[p]
\centering
\subfloat[]{\includegraphics[scale=0.82]{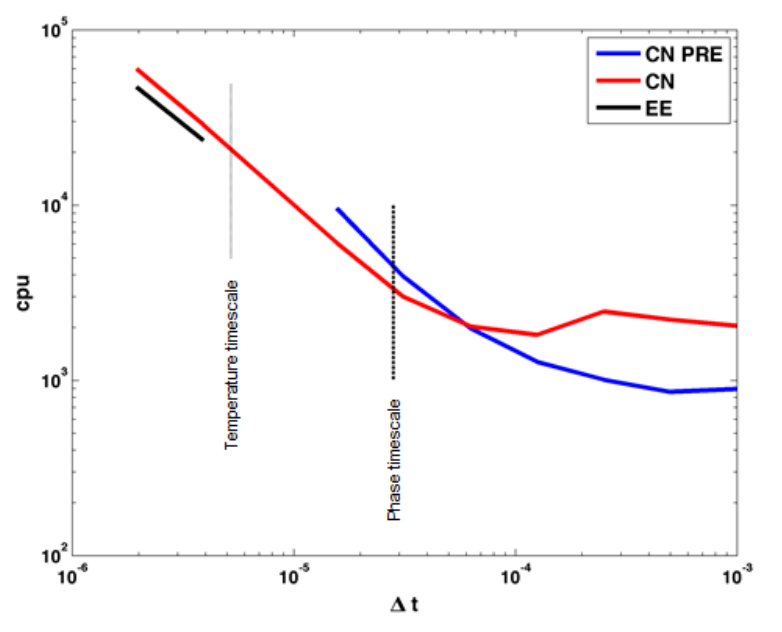}} \\
\subfloat[]{\includegraphics[scale=0.82]{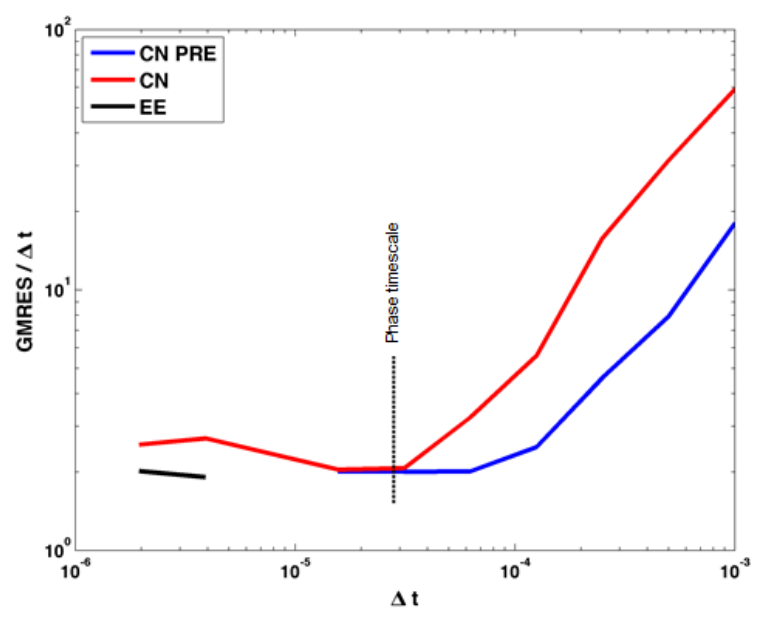}} \\
\subfloat[]{\includegraphics[scale=0.38]{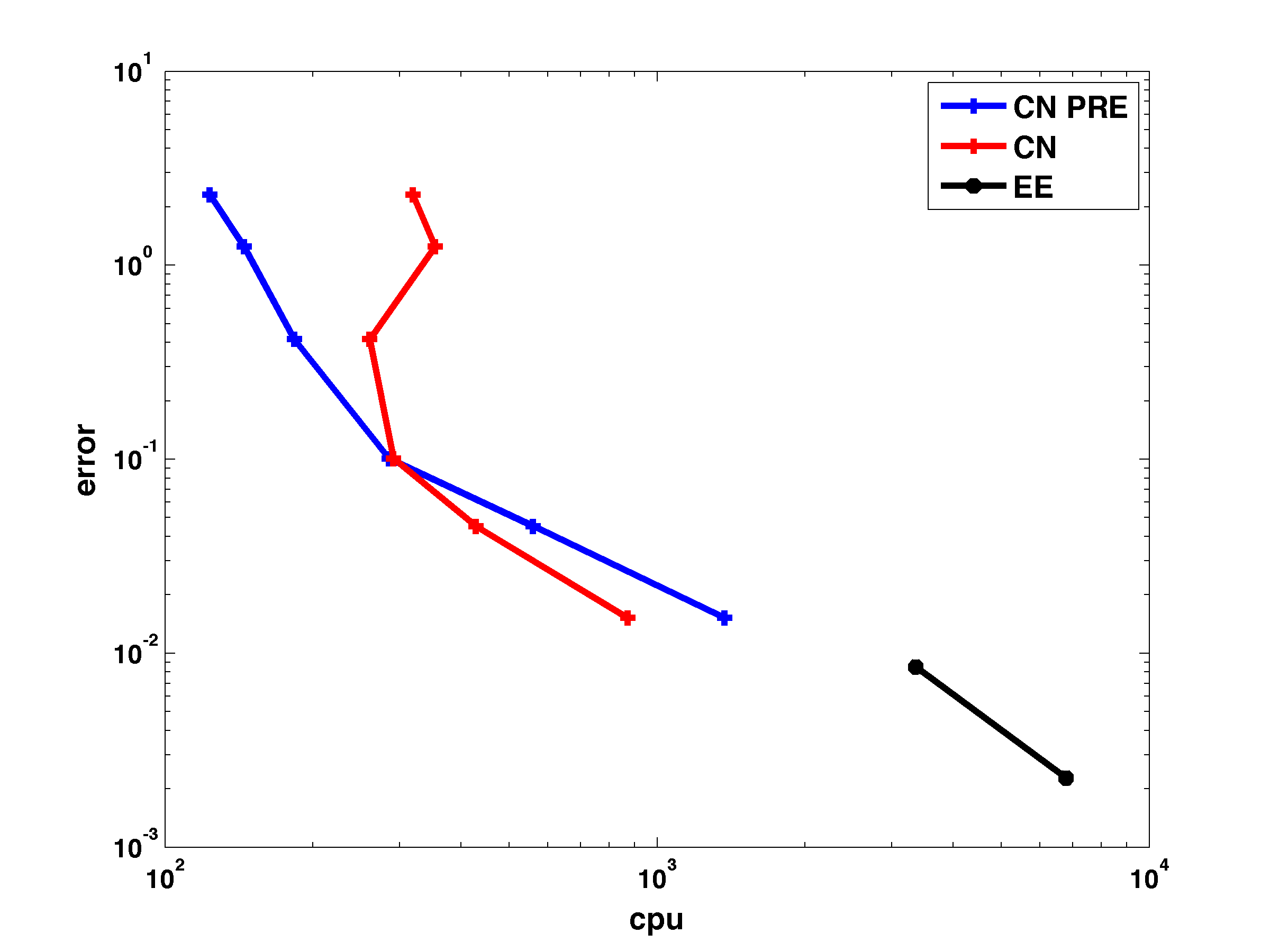}}
\caption{(a) Cost in terms of CPU time as a function of timestep size for explicit Euler (EE), Crank-Nicolson (CN) and Crank-Nicolson with preconditioning (CN-PRE) and reference lines for $\dt = \Dt_\temp$ (Temperature timescale) and $\dt = \Dt_\pf$ (Phase timescale). (b) Average number of GMRES iterations per timestep as a function of timestep size. (c) Error as a function of CPU time.}
\label{fig_preconditioner}
\end{figure}

To assess algorithmic performance of our method, we consider CPU time required for time integration to a fixed $t = t_f$ as a function of timestep size, $\Delta t$. 
We recall from Sec.~\ref{sec:timescale} that $\Delta t$ and $h$ are explicitly coupled through
Eqs.~\eqref{eq:dtpf}--\eqref{eq:dtt}.
In this context, we define a method as ideally algorithmically scalable if CPU time required for time integration is inversely proportional to $\Delta t$. With this definition, explicit methods are inherently ideally algorithmically scalable. Figure~\ref{fig_preconditioner}a shows CPU time as a function of $\Delta t$ for explicit Euler (EE), Crank-Nicolson (CN), and Crank-Nicolson with preconditioning (CN-PRE). Figure~\ref{fig_preconditioner}a confirms algorithmic scalability for EE for $\dt < \Dt_\temp$ (the explicit CFL limit). Algorithmic scalability is also confirmed for (unpreconditioned) CN for $\dt < \Dt_\pf$; however for $\dt > \Dt_\pf$, the cost of the method increases. When effectively preconditioned, CN scales well algorithmically for $\Delta t$ up to $4\,\Dt_\pf$.

Recall that the dominant cost in JFNK is GMRES iterations, and the key idea for preconditioning is to reduce the number of GMRES iterations per nonlinear solve for a desired level of accuracy for time integration independent of timestep size.
Figure~\ref{fig_preconditioner}b demonstrates the effect of preconditioning and shows the average number of GMRES iterations as a function of timestep size. The average number of GMRES iterations remains constant for CN up to $\dt=\Dt_\pf$ and increases rapidly for $\dt>\Dt_\pf$. In contrast, the average number of GMRES iterations remains constant for CN-PRE up to $\dt=4\,\Dt_\pf$. Additionally, our preconditioning approach reduces the number of GMRES iterations at each Newton step by a factor of 2 to 3. As a well-conditioned, non-symmetric mass matrix must be approximately inverted each timestep for EE, the average number of GMRES iterations per timestep for EE is also shown in Fig.~\ref{fig_preconditioner}b. 

Figure~\ref{fig_preconditioner}c shows error as a function of cost in terms of CPU time for EE, CN, and CN-PRE. Figure~\ref{fig_preconditioner}c shows EE to have less error, however costs considerably more than CN; and demonstrates that for a given level of accuracy, implicit methods can require less cost in terms of CPU time. In particular, for a given level of accuracy, CN-PRE can have less cost than CN. Effective preconditioning allows the implicit method to scale algorithmically with larger timesteps than explicit methods, while maintaining second-order temporal accuracy. 

\subsection{Directional growth}\label{sec:alloy}
Similar to free growth, directional crystallization of alloys is also a well-understood benchmark problem but more relevant for additive manufacturing~\cite{dantzigbook,kurzbook,francois2017,ghosh_review}. In this case, the evolution equations in Sec.~\ref{sec:metal} were modified to accommodate the evolving $\conc(\bx,t)$ and $\temp(\bx,t)$ fields. We model directional growth of a sample alloy in 2D and 3D following~\cite{Echebarria2004}. With a particular choice of $f(\pf, \conc)$, Eq.~\eqref{eq:dpf} becomes
\begin{multline}\label{eq:pf2}
F_1(\pfh^{n+1},\conch^{n+1})_i=
    \frac{1}{\dt}\left[\left(\left[ 1+(1-k)\conch^{n+1}\right] g_s^2(\atheta^{n+1}) {\pfh}^{n+1}, \psi_i\right) 
        - \left( \left[1+(1-k)\conch^{n}\right] g_s^2(\atheta^{n}) {\pfh}^{n}, \psi_i\right)\right]\\
    +\thetam\, \left(g_s^2(\atheta^{n+1}) \nabla\pf_h^{n+1}, \nabla \psi_i\right)
    +(1-\thetam) \left(g_s^2(\atheta^{n}) \nabla\pf_h^{n}, \nabla \psi_i\right)\\
    +\thetam\,\left(g_s(\atheta^{n+1})||\nabla \pf_h^{n+1}||^2, \nabla \psi_i\right)
    +(1-\thetam) \left(g_s(\atheta^{n})||\nabla \pf_h^{n}||^2, \nabla \psi_i\right)\\
    -\thetam\, \left(G_3(\pf_h^{n+1})(\conch^{n+1}+G_4), \psi_i\right)
    -(1-\thetam) \left(G_3(\pf_h^{n})(\conch^{n}+G_4), \psi_i\right),
\end{multline}
with 
\begin{equation}\label{eq:lambda}
    G_3(\pfh)=\pf_h - \pf_h^3 - \lambda ( 1-\pf_h^2 )^2
\end{equation}
and
\begin{equation}\label{eq:directional}
    G_4(x, t) = \frac{G(x-R\,t)}{|m|c_\infty(1-k)/k},
\end{equation}
where $x$ is the direction of thermal gradient or dendrite growth. Similarly, Eq.~\eqref{eq:dc} for $u$ becomes
\begin{multline}\label{eq:conc2}
F_2(\pf_h^{n+1},\conc_h^{n+1})_i=
    \frac{1}{\dt}\left[\left(\frac{1+k}{2} \conch^{n+1}, \psi_i\right) 
        - \left(\frac{1+k}{2} \conch^n, \psi_i\right)\right]\\
  +\thetam\,\left(\frac{1-\pfh^{n+1}}{2}\,D\, \nabla\conch^{n+1}, \nabla\psi_i\right)
   +(1-\thetam)\left(\frac{1-\pfh^{n}}{2} \,D\, \nabla\conch^{n}, \nabla\psi_i\right) \\
  + \left(\jat(\widehat{\pfh^n}), \nabla\psi_i\right) 
  - \left(\frac{1}{2} \widehat{\pfh} \left[1+(1-k)\conch^{n+1}\right], \psi_i\right),
\end{multline}
with $\widehat{\pf_h}$ given by Eq.~\eqref{eq:theta}. 
We note that Eqs.~\eqref{eq:pf2}--\eqref{eq:conc2} are nondimensional, where the constant $\lambda$ in Eq.~\eqref{eq:lambda} characterizes the length scale of the interface $W_0 = d_0 \lambda/a_1$ and the time scale $\tau_0 = (d_0^2/D_L)a_2\lambda^3/a_1^2$, with $a_1 = 0.8839$, $a_2 = 0.6267$ and $D=D_L \tau_0/W_0^2$. Nondimensionalization assures that $\pf=\mathcal{O}(1)$ and $\conc=\mathcal{O}(1)$, thus balancing the magnitudes of the variables and residuals and allowing convergence in the Newton solve. Furthermore, the length scale $W_0$ is chosen such that we obtain converged quantitative results independent of grid spacing $h/W_0$~\cite{Echebarria2004,farzadi:2008}. Our phase-field simulations are fully converged and the diffuse interface is fully resolved for $h = 0.8$ and no further increase in mesh resolution is required.
The anti-trapping current, 
\begin{equation}\label{eq:jat}
    \jat(\widehat{\pfh}) = \frac{1}{2\sqrt{2}} \left[ 1+(1-k)\conch\right] \widehat{\pfh} \nabla \pfh,
\end{equation}
avoids artificial interfacial effects on solute field~\cite{karma2001}. The nondimensional concentration $\conc$ is related to solute composition $c$ (wt\%) by
\begin{equation}\label{eq:u}
%\conc=\ln \left(\frac{2\ c\ k/c_\infty}{1+k-(1-k)\varphi}\right).
\conc = \frac{\left(\frac{2\ c\ k/c_\infty}{1+k-(1-k)\varphi}\right)-1}{1-k}.
\end{equation}
Material properties and numerical parameters are listed in Table~\ref{tab:ds_parameters}.
We utilize 9 quadrature points in each quadrilateral element (2D) and 27 quadrature points in each hexahedral element (3D) to accurately integrate the nonlinear terms in Eqs.~\eqref{eq:pf2}--\eqref{eq:jat}.

\begin{table}[!t]
  \begin{centering}\footnotesize
    \begin{tabular}{ccccccccc}\hline
      $\epsilon$&Anisotropy& $0.01$&&&&
      %$M$&lobe number& $4.0$\\
      $k$&Equilibrium partition coefficient& $0.14$ \\
      $m$ & Liquidus slope (K/wt\%) & $-2.6$ &&&&
      $c_\infty$&Alloy (Al-Cu) composition (wt\%)&$3$\\
      $\lambda$&Coupling constant& $10.0$&&&&
      $D_L$ & Liquid diffusivity (m$^2$/s) & $3 \times 10^{-9}$\\
      $d_0$ & Capillary length (m) &$5\times10^{-9}$&&&&
      $G$ & Thermal gradient (K/m) &  $10^4$ \\
      $\Gamma$ & Gibbs-Thomson coefficient (K m) & $2.4 \times 10^{-7}$ &&&&
      $R$ & Growth rate (m/s) & $10^{-2}$
\\ \hline
    \end{tabular}
    \caption{Material properties and simulation parameters in Eqs.~\eqref{eq:pf2}--\eqref{eq:u}, adopted from~\cite{farzadi:2008}.}
    \label{tab:ds_parameters}
  \end{centering}
\end{table}

Simulations were performed on meshes consisting of $1024\times256$ (2D) and $1024\times256\times256$ (3D) elements with $h = 0.8$, implicit time integration with $\thetam = .5$ and $\Delta t = 0.002$, and zero Neumann boundary conditions on $\partial\Of$. Initial conditions consist of a solid-liquid interface at $x=8$ units, with 
\begin{equation}
\pf(\bx,0)= \left\{
    \begin{array}{rcc} 
    1&:& 0\le x\le 8\\
    -1&:& x> 8 
    \end{array} 
    \right.
    ,
\end{equation}
where $\pf = 1$ corresponds to solid and $\pf = -1$ corresponds to liquid. 
The initial solute concentration $u(\bx,0) = -1$ corresponds to $c = k\ c_\infty$ in solid and $c = c_\infty$ in liquid. A small, random perturbation is applied to the initial solid-liquid interface.

\begin{figure}[t]
\centering
\subfloat[]{\includegraphics[scale=0.51]{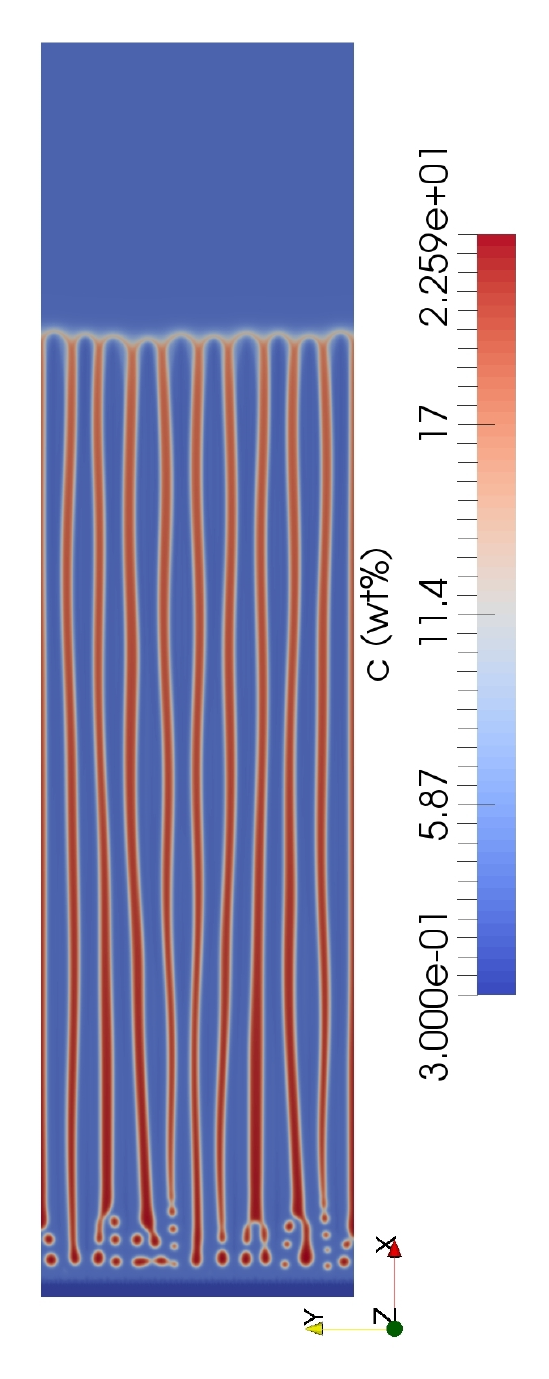}}\hspace{30mm}
\subfloat[]{\includegraphics[scale=0.352]{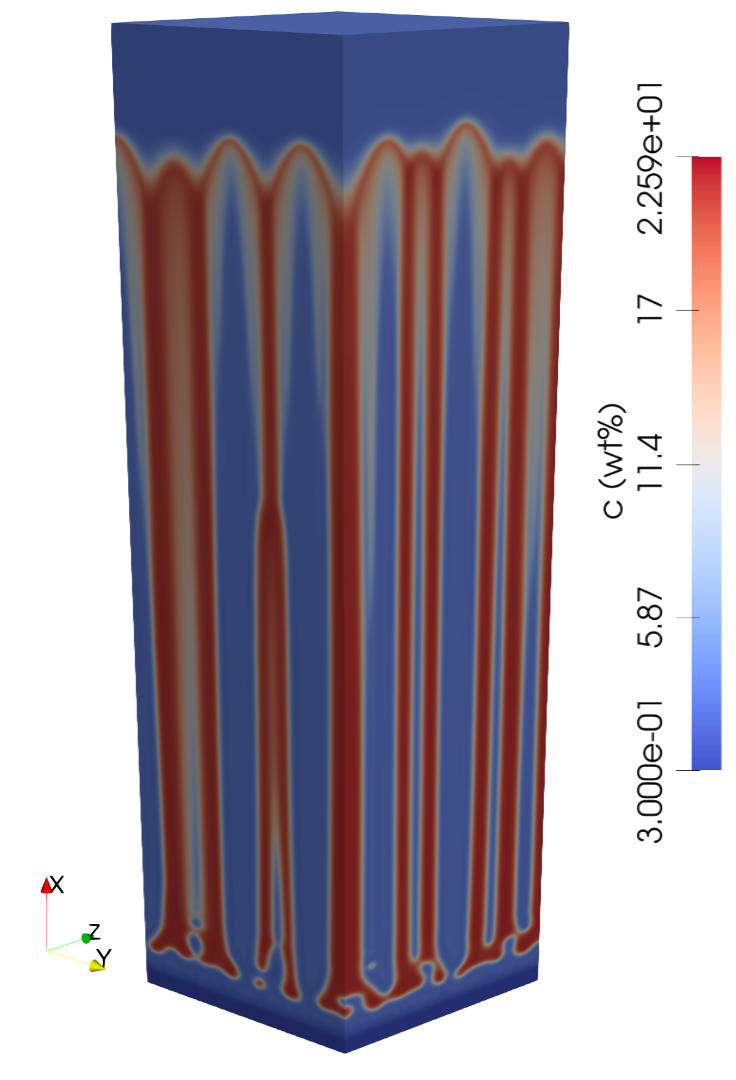}}
\caption{Alloy dendrite evolution using the phase-field model (Eqs.~\eqref{eq:pf2}--\eqref{eq:conc2}) and physical properties (Table~\ref{tab:ds_parameters}). (a) 2D simulation snapshot for dimensionless $t_f = 170$. The composition field (wt\%) is shown at the steady-state. The dimensional solute redistribution profile is shown using a color bar. (b) The corresponding 3D simulation snapshot. The gradient or growth direction, $x$, is vertical in the figure.}
\label{fig_farzadi}
\end{figure}

The perturbation grows with time, starting with the transient stages of growth followed by steady-state, when the dendrite tips grow with constant velocity. The evolution of columnar dendritic microstructure is shown in 2D in Fig.~\ref{fig_farzadi}a, and the same is shown in 3D in Fig.~\ref{fig_farzadi}b. The finger-like cellular features are colored by $c$. Simulations are essential to study materials behavior in applications, for example, the redistribution of $c$ across, along, and through the cells determine the solute-trapping behavior of the alloy material~\cite{supriyo20173d,ghosh_msmse2017,Ghosh2018_scripta,farzadi:2008}. This can be easily explored in detail by varying $G$ and $R$ in our simulations using the \tusas\ framework; a study of which is beyond the scope of this work.

The 3D simulations were performed on \texttt{Summit} at Oak Ridge National Laboratory~\cite{summit} across multiple GPUs and multiple nodes using MPI and CUDA. 
The 3D simulations demonstrate the capability of \tusas\ to perform large-scale long-time simulations using next-generation parallel computing techniques; the details of our hybrid parallel performance is given in Sec.~\ref{sec:performance}.

\begin{figure}[t]
\centering
\subfloat[]{\includegraphics[scale=0.35]{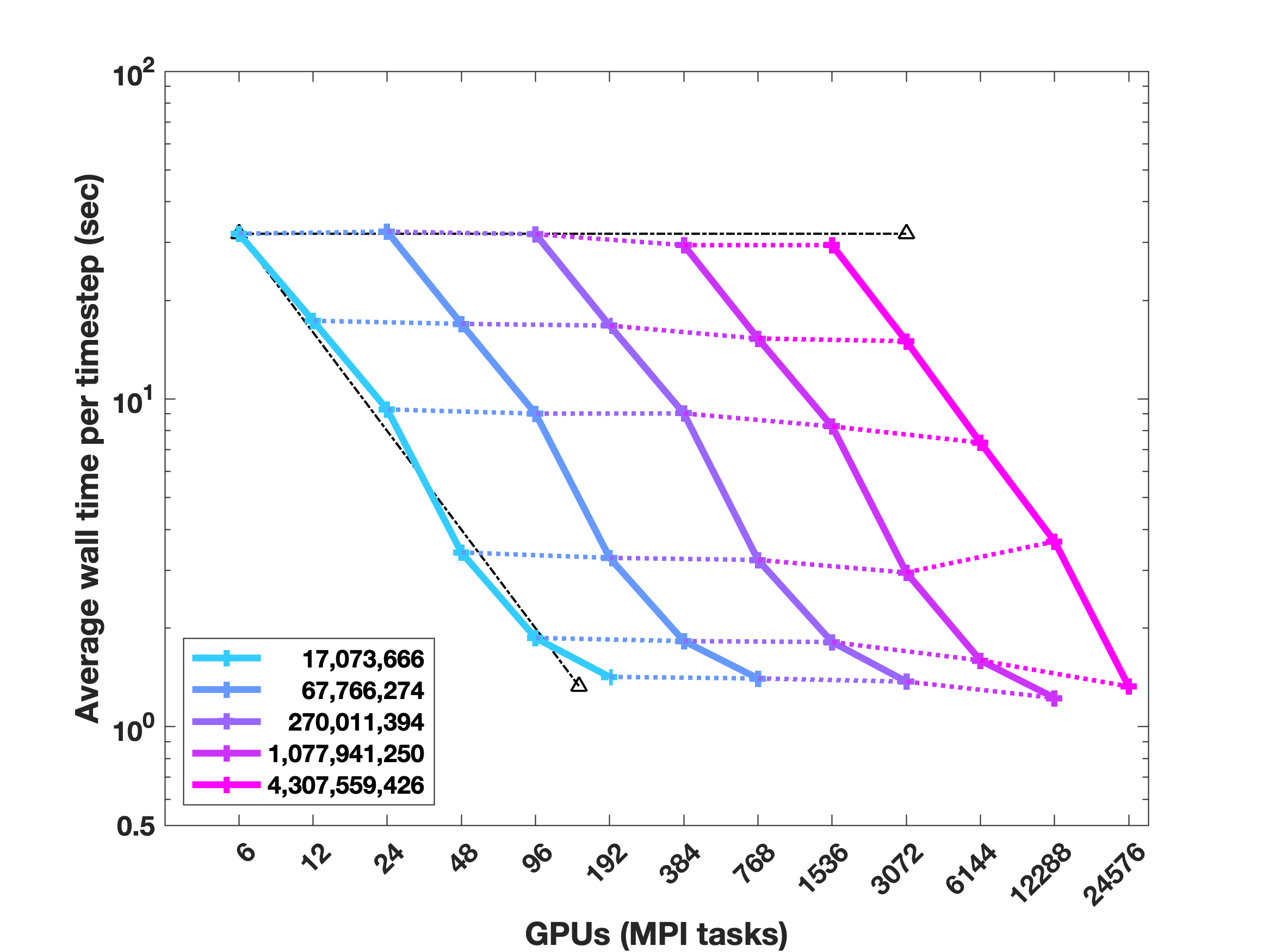}}%
\subfloat[]{\includegraphics[scale=0.35]{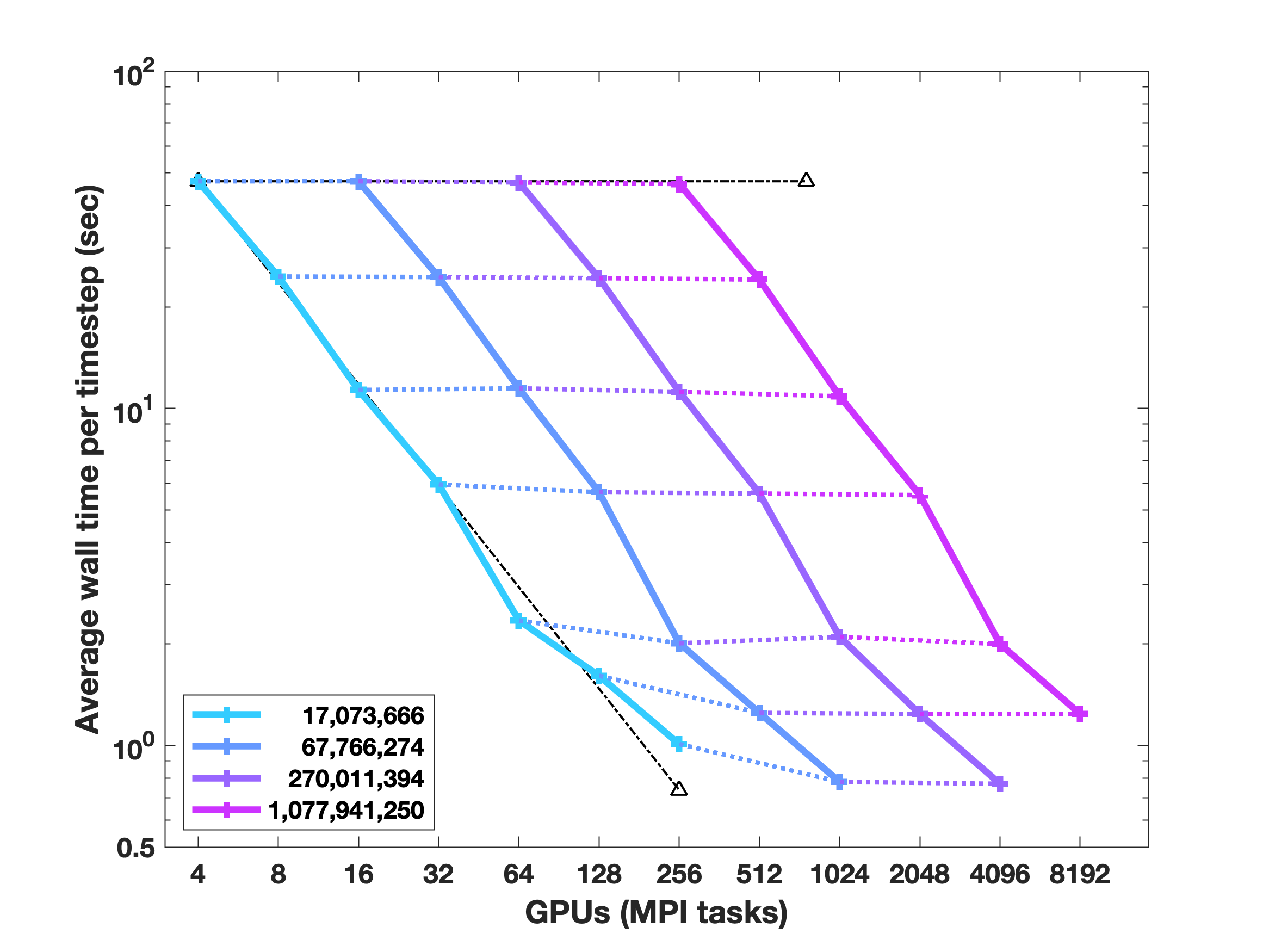}}
\caption{(a) Strong and weak multi-node, multi-GPU scaling on \summit. (b) Strong and weak multi-node, multi-GPU scaling on \sierra.}
\label{fig_performance}
\end{figure}

\subsection{Parallel performance}\label{sec:performance}
In addition to performing well in parallel across multiple CPUs using MPI and OpenMP, a particular strength of \tusas\ is the ability to run efficiently across multiple GPUs using MPI and CUDA. As the residual fill in Eqs.~\eqref{eq:pf2}--\eqref{eq:conc2} is composed entirely of local operations, it is performed entirely on the GPU utilizing a distance-1 element graph coloring within each MPI subdomain. Element graph coloring avoids shared memory race conditions and atomic operations on CPU threads and GPUs. In addition, inner products within the GMRES solver are implemented on the GPU \textit{via} the \texttt{Belos} and \texttt{Kokkos} packages in \texttt{Trilinos}~\cite{trilinos}.

To demonstrate strong and weak parallel multi-node and multi-GPU scaling, we consider simulations for the 3D problem discussed in Sec.~\ref{sec:alloy} with discretizations consisting of $17\,073\,666$, $67\,766\,274$, $270\,011\,394$, $1\,077\,941\,250$ and $4\,307\,559\,426$ unknowns. The discretizations consist of meshes with $512\times 128\times 128$, $512\times 256\times 256$, $512\times 512\times 512$, $512\times 1024\times 1024$ and $512\times 2048\times 2048$ bilinear hexahedral elements. 

Scaling results were performed on \texttt{Summit} at Oak Ridge National Laboratory~\cite{summit} and \texttt{Sierra} at Lawrence Livermore National Laboratory~\cite{sierra}. \texttt{Summit} is comprised of $4,608$ nodes, each node consists of 2 IBM Power9 CPUs, 6 NVIDIA Volta V100 GPUs and 512 GB memory, and achieves $148.6$ PetaFLOPS peak performance; \texttt{Sierra} is comprised of $4,320$ nodes, each node consists of 2 IBM Power9 CPUs, 4 NVIDIA Volta V100 GPUs and 256 GB memory, and achieves $94.6$ PetaFLOPS peak performance.

Strong scaling is defined by how CPU time varies with the number of processors for a fixed total problem size and is used to determine if the software is compute-bound. Ideal strong scaling is achieved if the CPU time is inversely proportional to the number of processing elements used. Weak scaling is defined by how CPU time varies with the number of processors for a fixed problem size per processor and is used to determine if the software is memory-bound. Ideal weak scaling is achieved if the CPU time remains constant while the workload is increased in direct proportion to the number of processors.

Figure~\ref{fig_performance} shows the CPU time required for time integration to a fixed time as a function of the number of GPUs. Specifically, Fig.~\ref{fig_performance}a shows strong and weak scaling on \texttt{Summit} on up to over 4 billion unknowns on up to $24,576$ GPUs ($4,096$ nodes), and Fig.~\ref{fig_performance}b shows strong and weak scaling on \texttt{Sierra} on up to over 1 billion unknowns on up to $8,192$ GPUs ($2,048$ nodes). Strong scaling is depicted by solid lines on Fig.~\ref{fig_performance}, where each color depicts a fixed problem size. Weak scaling is depicted by horizontal markers. This preliminary scaling study demonstrates that \tusas\ effectively scales strongly and weakly on thousands of GPUs with billions of unknowns and is particularly suited for emerging heterogeneous architectures~\cite{summit,sierra,shimokawabe2011}. A detailed study of parallel implementation, performance, scalability, and efficiency is beyond the scope of this work and will be the focus of a separate manuscript.

\newcommand{\cfl}{\kappa}
\subsection{Algorithmic performance of our preconditioning strategy} 
We consider the following linearized forms of Eq.~\eqref{eq:pf2} and Eq.~\eqref{eq:conc2} given by:
\begin{equation}\label{eq:pref2}
(M_{11})_{ij}=
    \frac{1}{\dt}\left(\left[ 1+(1-k)\conch^{n+1}\right] g_s^2(\atheta^{n+1}) \psi_j, \psi_i\right)
    +\thetam\, \left(g_s^2(\atheta^{n+1}) \nabla\psi_j, \nabla \psi_i\right),
\end{equation}
and 
\begin{equation}\label{eq:preconc2}
(M_{22})_{ij}=
    \frac{1}{\dt}\left(\frac{1+k}{2} \psi_j, \psi_i\right)
  +\thetam\,\left(\frac{1-\pfh^{n+1}}{2}\,D\, \nabla\psi_j, \nabla\psi_i\right),
\end{equation}
respectively.
We consider the dominant timescale for Eq.~\eqref{eq:conc2}, given by
\begin{equation}
    \Dt_\conc =\frac{h^2}{2 D_0},
\end{equation}
where $D_0=D/(1+k)$ and define $\cfl(\Delta t,h)=2 \Delta t D_0 / h^2$.
We note that $\cfl=1$ represents the (scaled) timestep limit for explicit time integration.
We recall that the goal is to capture the stiff, elliptic terms of Eqs.~\eqref{eq:pf2}--\eqref{eq:conc2} in our preconditioning operator to allow algorithmically scalable time integration for $\cfl>1$, by reducing the number of GMRES iterations.
Hence, for implicit time integration without preconditioning, we expect good performance and algorithmic scaling of the solver for $\cfl<1$ and deteriorating performance for $\cfl>1$; and expect the preconditioned solver to scale beyond $\cfl=1$.

There has been many recent developments of algebraic multigrid using GPUs which include both classical and aggregation-based methods ~\cite{liu2015amg,haase2010amg,richter2014amg,gandham2014amg,MueLu,MueLuURL}.
For this example, we utilize the \texttt{MueLu} multigrid package~\cite{MueLu,MueLuURL} within the \texttt{Trilinos} 12.18 library with unsmoothed aggregation, a single V-cycle with 5 levels and degree-2 Chebyshev smoothing on each level, and a maximum of 4 levels. \texttt{MueLu} has shown to be scalable across CPUs, threads and GPUs~\cite{tuminaro2019}. Results were obtained on \texttt{Summit}~\cite{summit}. For this example, both the residual and preconditioner operator fills are performed on the GPUs; additionally, the associated multigrid algorithms are performed on the GPUs via the \texttt{MueLu} library.
We consider CPU time required for time integration to fixed time $t_f=10$ as a function of $\cfl$ with $\cfl=0.1678$, 0.2684, 0.3355, 0.6711, 1.6777, 2.3488, 2.6844 and 3.3555.

\begin{figure}[t]
\centering
\subfloat[]{\includegraphics[scale=0.35]{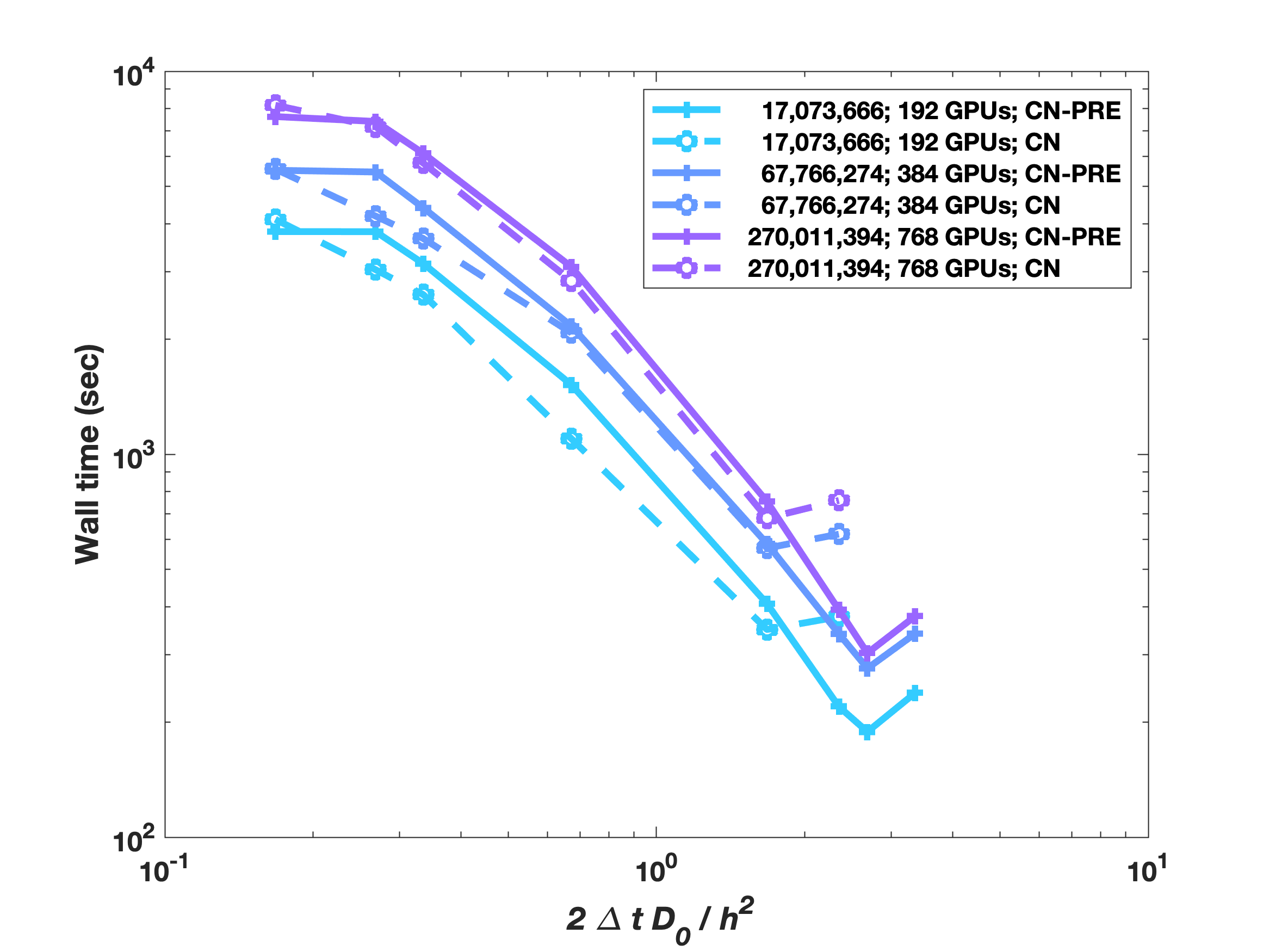}}%
\subfloat[]{\includegraphics[scale=0.35]{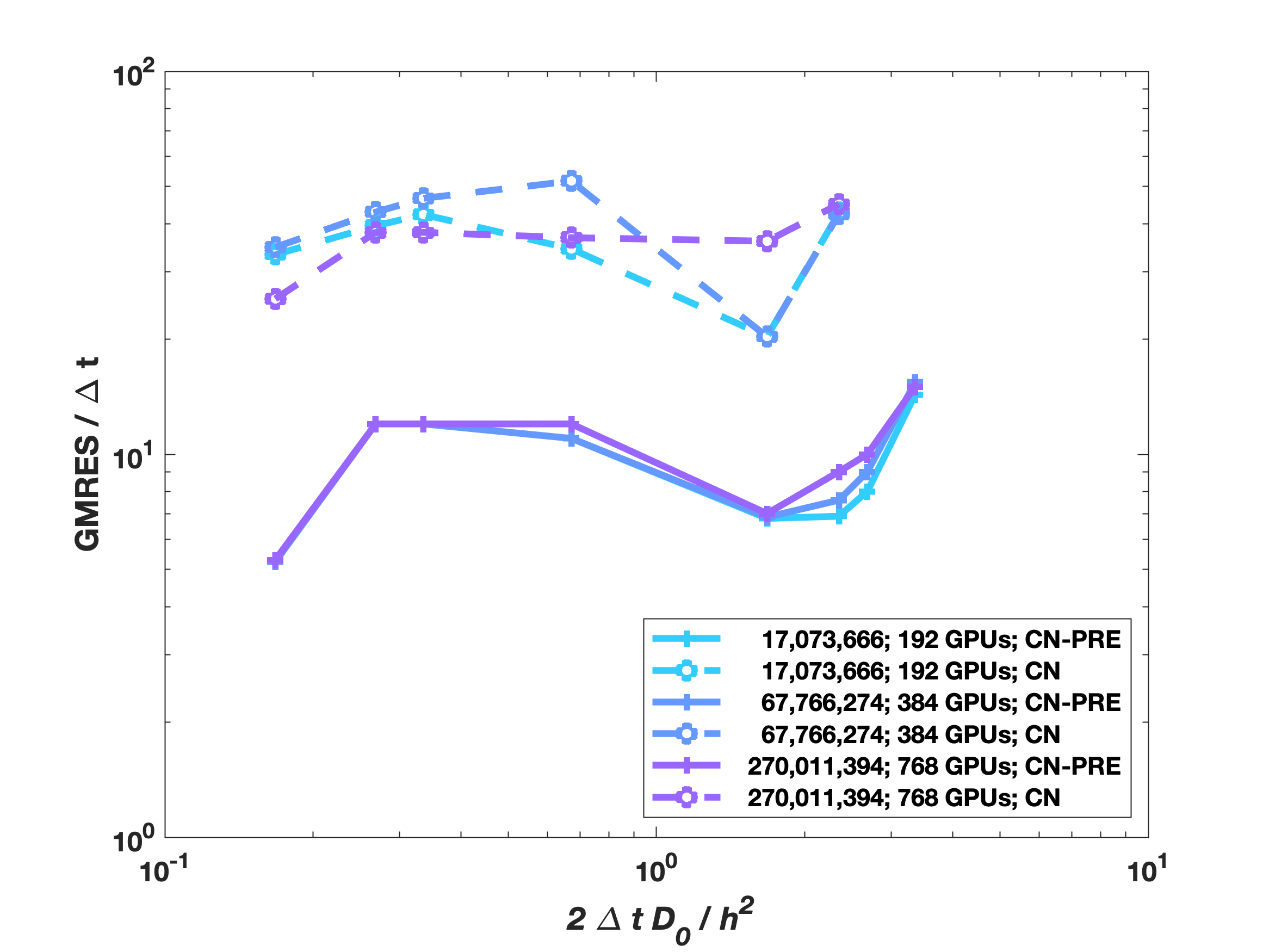}}
\caption{(a) Cost in terms of CPU time as a function of $\cfl$ for Crank-Nicolson (CN) and Crank-Nicolson with preconditioning (CN-PRE) for $17\,073\,666$ unknowns on 192 GPUs (32 nodes),
$67\,766\,274$ unknowns on 384 GPUs (64 nodes) and $270\,011\,394$ unknowns on 768 GPUs (128 nodes). 
(b) Average number of GMRES iterations per timestep as a function of $\cfl$.}
\label{fig_preconsummit}
\end{figure}

Figure~\ref{fig_preconsummit} demonstrates the effect of preconditioning for this particular choice of multigrid parameters and shows performance as a function of $\cfl$ for Crank-Nicolson (CN) with $\thetam=.5$, and Crank-Nicolson with preconditioning (CN-PRE) for the following cases:
\begin{enumerate}
\item $17\,073\,666$ unknowns on 192 GPUs (32 nodes),
\item $67\,766\,274$ unknowns on 384 GPUs (64 nodes),
\item $270\,011\,394$ unknowns on 768 GPUs (128 nodes),
\end{enumerate}
and shows CN as dashed lines, CN-PRE as solid lines; problem size is depicted by color.
Figure~\ref{fig_preconsummit}a shows CPU time as a function of $\cfl$,
and confirms algorithmic scalability for CN for $\cfl<1$; however for $\cfl>1$ the cost of the method increases. When preconditioned, CN-PRE scales algorithmically up to $2.6\times\cfl$.
Both CN and CN-PRE scale algorithmically, independent of both problem size and number of GPUs,
with CN-PRE costing negligibly more do to the overhead of the algebraic multigrid algorithm.
Figure~\ref{fig_preconsummit}a confirms strong scaling of the preconditioner on up to 270M unknowns on up to 768 GPUs.
Currently, with this particular choice of multigrid parameters, we see a decrease in cost by a factor of two using our preconditioning strategy.  

Figure~\ref{fig_preconsummit}b shows the average number of GMRES iterations per timestep as a function of $\cfl$. The average number of GMRES iterations remains constant for both CN and CN-PRE up to $\cfl\approx 1$, with preconditioning reducing the average number of GMRES iterations by a factor of up to four. Figure~\ref{fig_preconsummit}b confirms, with this particular choice of multigrid parameters, the preconditioning reduces the average number of GMRES iterations to $2.6\times\cfl$, independent of both problem size and number of GPUs.

In the present example, we have evaluated our preconditioning strategy with a particular choice of algebraic multigrid parameters as suggested in~\cite{MueLu}, and a particular choice of linearization of the Jacobian operator. There is a numerous and diverse set of algebraic multigrid parameters applicable to this problem~\cite{MueLu}. Additionally, there are many choices related to preconditioner operator construction, including linearization, block-Jacobi iteration, and automatic differentiation approaches. This example serves as an initial study to demonstrate the potential efficacy of our preconditioning approach. A detailed analysis is beyond the scope of this effort and will be the focus of a separate manuscript.

%----------------------------------------------------------------------------------

\section{Discussion and summary}\label{sec:summary}
We have presented an implicit approach for predictive phase-field simulation of solidification. The method is based on implicit time integration, where the solution of the nonlinear system is approximated by JFNK with physics-based preconditioning. Spatial discretization consists of an unstructured, nodal finite element method. Such a combination of advanced, modern numerical approaches and preconditioning facilitates ideal algorithmic efficiency and scalability at timesteps greater than that traditional CFL limits allow.

In this edition, we introduce a new open-source phase-field modeling framework, \tusas, and demonstrate its efficacy on phase-field simulations of dendritic solidification for pure and binary alloy materials assuming both isothermal and directional growth conditions. The framework is constructed using modern computational science and solver libraries and is implemented utilizing a hybrid parallel approach for CPU/GPU architectures. We have demonstrated ideal strong and weak scaling on up to 4 billion unknowns and thousands of GPUs using benchmark phase-field simulations on \summit~\cite{summit} and \sierra~\cite{sierra}. These scaling studies signify that \tusas\ is well-suited for next-generation high-performance supercomputer architectures.

Currently, \tusas\ is playing a critical role in studying the solidification behavior of metals and alloys in engineering processes, in particular, additive manufacturing~\cite{francois2017,farzadi:2008}, as we continue to prepare for exascale computing~\cite{kothe2018exascale,exascale2020}. In conclusion, we have shown that modern algorithms, discretizations, and computational science, and heterogeneous hardware provide a robust route for predictive phase-field simulation of microstructures that will complement experimental studies and provide significant contributions to better understanding material behavior rapidly.

%%------------------------------------------------------------------------------------------------------
\section*{Declaration of competing interest}
The authors declare that they have no known competing financial interests or personal relationships that could have appeared to influence the work reported in this paper.

\section*{Acknowledgments}
This work was supported by the U.S. Department of Energy through the Los Alamos National Laboratory. Los Alamos National Laboratory is operated by Triad National Security, LLC, for the National Nuclear Security Administration of the U.S. Department of Energy (Contract No.\linebreak 89233218CNA000001) and by the Exascale Computing Project (17-SC-20-SC), a joint project of the U.S. Department of Energy's Office of Science and National Nuclear Security Administration, responsible for delivering a capable exascale ecosystem, including software, applications, and hardware technology, to support the nation's exascale computing imperative (LA-UR-20-23977).

%\section*{References}
%\bibliography{bibtex/tusas,bibtex/localabbrev,bibtex/mrabbrev,bibtex/external,bibtex/mmg_journals}
%\nocite{*}

%--------------------------------------------------
\end{document}